\documentclass[10pt,aps,prd,twocolumn,showpacs,superscriptaddress,longbibliography,nofootinbib]{revtex4-2}
\usepackage{mathrsfs,amsmath,amsthm,latexsym,amssymb,amsfonts,epsfig,cancel,enumerate,graphicx,txfonts,diagbox}
\usepackage{tikz}
\usepackage{float}
\usepackage{flushend}
\usepackage{colortbl}
\usepackage{flushend}
\usepackage{multirow}
\usepackage[T1]{fontenc}
\usepackage[utf8]{inputenc}
\usepackage[colorlinks=true,linkcolor=blue,citecolor=blue,urlcolor=blue]{hyperref}
\usepackage{orcidlink}

\usepackage{xcolor}
\usepackage{booktabs}
\usepackage{graphicx}
\usepackage{subcaption}

\definecolor{navy}{RGB}{0,0,150}
\usepackage{appendix}
\allowdisplaybreaks
\newcommand{\RGU}{Department of Physics, The Assam Royal Global University, Guwahati-781035, Assam, India}
\newcommand{\ABTU}{Department of Physics, Al-Hussein Bin Talal University, 71111, Ma'an, Jordan}
\newcommand{\EMU}{Physics Department, Eastern Mediterranean University, Famagusta 99628, North Cyprus via Mersin 10, Turkey}

\begin{document}

\title{Shadow, Sparsity of Radiation and Energy Emission Rate in Skyrmion Black Holes}

\author{Faizuddin Ahmed\orcidlink{0000-0003-2196-9622}}
\email{faizuddinahmed15@gmail.com}
\affiliation{\RGU}

\author{Ahmad Al-Badawi\orcidlink{0000-0002-3127-3453}}
\email{ahmadbadawi@ahu.edu.jo}
\affiliation{\ABTU}

\author{\.{I}zzet Sakall{\i}\orcidlink{0000-0001-7827-9476}}
\email{izzet.sakalli@emu.edu.tr (Corresp. author)}
\affiliation{\EMU}

\begin{abstract}
We examine several observable optical properties of a Skyrmion black hole (BH), focusing on the photon sphere, BH shadow, and photon trajectories. The Skyrme term, along with other geometric parameters of the spacetime, determines the photon sphere location and shapes the resulting BH shadow. Parameter variations produce observable departures from standard BH geometries, offering potential signatures of nonlinear field effects. We also analyze the sparsity of Hawking radiation and the associated energy emission spectra, showing how these quantities respond to the Skyrme coupling and background parameters. Our findings illuminate the connection between nonlinear field contributions and BH optics, with implications for observational and theoretical studies of modified gravity scenarios.
\end{abstract}

\maketitle

\section{Introduction}\label{sec:1}

BH physics has gained renewed attention following the first gravitational wave detection~\cite{Abbott2016,Abbott2017} and the groundbreaking shadow observations by the Event Horizon Telescope (EHT)~\cite{EHTL1,EHTL12}. Information encoded in BH images deepens our understanding of shadows, jets, and accretion processes. These achievements confirm Einstein's general relativity (GR) in the strong-field regime while advancing BH physics and opening pathways to test GR limits, potentially uncovering deviations or extensions to Einstein's theory.

Current BH images do not yet permit precise geometric identification, though observational improvements will deliver higher resolution in coming years. Theoretical work predicting BH shadow shapes across various gravity theories and astrophysical settings remains essential. The BH shadow appears as a two-dimensional dark region to distant observers when a bright background source lies behind it. Shadow studies yield valuable BH property information and serve as powerful GR tests. BH shadows have become central to modern research, particularly for interpreting upcoming high-resolution data. Shadow investigations across different BH spacetimes appear extensively in the literature~\cite{Volker2022}.

BH shadow research began with Synge~\cite{Synge1966}, who examined the Schwarzschild BH shadow. Bardeen~\cite{Bardeen1973} subsequently analyzed the rotating Kerr BH shadow, while Luminet~\cite{Luminet1979} studied thin accretion disk effects on shadow formation. Direct imaging of supermassive BHs M87$^*$ and Sgr~A$^*$ by the EHT collaboration~\cite{EHTL1,EHTL12} has intensified shadow research interest. These observations opened a new window into the strong-gravity regime. BH internal structure cannot be probed directly due to their defining properties, yet interactions with the surrounding environment through scattering, absorption, and Hawking radiation provide indirect insights.

Nonlinear sigma models rank among the most important nonlinear field theory classes given their broad applications spanning quantum field theory to statistical mechanics~\cite{Manton2007}. Examples include quantum magnetism, the quantum Hall effect, meson dynamics, and string theory. These models also serve as effective field theories, describing superfluid $^3$He among other systems.

Nonlinear sigma models in $(3+1)$ dimensions lack static soliton solutions with finite energy, as scaling arguments demonstrate. Skyrme overcame this limitation by introducing an additional higher-derivative term, now called the Skyrme term, which stabilizes the theory and permits finite-energy static solutions known as Skyrmions~\cite{Skyrme1961a,Skyrme1961b,Skyrme1962}. Excitations around Skyrmion configurations can be interpreted as fermionic degrees of freedom, making them suitable nucleon models. The Skyrme model has consequently become foundational in nuclear and high-energy physics.

The coupled Einstein-Skyrme equations admit a static, spherically symmetric anti-de Sitter (AdS) BH solution~\cite{Skyrme1961a,Skyrme1961b,Skyrme1962,Canfora2013,Canfora2014}, described by the line element (in geometrized units)
\begin{equation}
    ds^2 = -f(r)\,dt^2 + \frac{dr^2}{f(r)} + r^2 (d\theta^2 + \sin^2{\theta}\,d\phi^2),\label{metric}
\end{equation}
where the lapse function reads
\begin{align}
     f(r)= 1 - 8\pi K - \frac{2 M}{r} + \frac{4\pi K \lambda}{r^{2}}.\label{function}
\end{align}
Here $M$ denotes the Nucamendi-Sudarsky mass, $K=F_\pi^2/4$ is the Skyrme coupling constant, and $\lambda=4/(e^2 F_\pi^2)$ is the quartic Skyrme parameter, where the positive coupling parameters $(F_{\pi}, e)$ are phenomenologically given by~\cite{Canfora2014,Adkins1983}
\begin{equation}
    F_{\pi}=0.141\ \textrm{GeV}, \qquad 5 \le e \le 7.\label{aa2}
\end{equation}

Setting $\lambda = 0$ in solution~(\ref{metric}) reduces it to a global monopole-like spacetime~\cite{Barriola1989}. The Skyrme term contributes to the metric function similarly to the Maxwell term in the Reissner-Nordstr\"om solution, though unlike the Maxwell case, the $1/r^2$ coefficient is not an integration constant but is determined by the theory's couplings. Skyrmion BH thermodynamic properties, with and without cosmological constant, have been studied extensively, including extended phase space thermodynamics~\cite{Daniel2019}, quantum-corrected thermodynamics and $P$-$V$ criticality~\cite{Yawar2020}, phase structure~\cite{David2026}, and deflection angle in the string-field limit~\cite{Canfora2018,Sucu:2026ijgmmp}. These works collectively illuminate the interplay between Skyrme field dynamics, BH geometry, and gravitational phenomena.

Motivated by EHT BH image releases, we examine observable optical characteristics within Einstein-Skyrme theory. The Skyrme coupling constant $K$ and quartic parameter $\lambda$ govern the underlying nonlinear field strength and configuration, leaving direct imprints on BH geometry and influencing optical features including the photon sphere, shadow size, and photon trajectories. Studying these effects enhances theoretical understanding of BH spacetimes in nonlinear field theories while providing experimental detection avenues through high-resolution astronomical observations. We also examine the sparsity parameter, a dimensionless quantity connecting BH geometric size with Hawking radiation, to explore how $K$ and $\lambda$ modify the radiation profile and particle emission characteristics. Finally, we compute the Hawking radiation energy emission rate, showing how Skyrme parameter variations alter the emission spectrum. These investigations reveal how nonlinear field dynamics produce measurable signatures in both optical and radiative BH properties, bridging theoretical predictions with observational prospects.

\section{Observable Properties of BH}

This section examines null geodesics, emphasizing circular orbits, the BH shadow, and photon trajectories around spacetime~(\ref{metric}).

The null geodesic condition $ds^2=0$ applied to metric~(\ref{metric}) yields
\begin{equation}
    -f(r) \dot{t}^2+\frac{1}{f(r)} \dot{r}^2+r^2 \dot{\theta}^2+r^2 \sin^2\theta\, \dot{\phi}^2=0,\label{bb1}
\end{equation}
where dots denote differentiation with respect to an affine parameter $\tau$.

The static, spherically symmetric spacetime admits two conserved quantities associated with temporal $t$ and azimuthal $\phi$ coordinates:
\begin{equation}
    \mathrm{E}=f(r) \dot{t}\quad,\quad \mathrm{L}=r^2 \sin^2\theta\, \dot{\phi},\label{bb2} 
\end{equation}
where $\mathrm{E}$ and $\mathrm{L}$ represent the conserved energy and angular momentum of test particles.

For null geodesic motion in the equatorial plane with $\theta=\pi/2$ and $\dot{\theta}=0$, the photon equation of motion becomes
\begin{equation}
    \dot{r}^2=\mathrm{E}^2-V_{\rm eff}(r),\label{bb3}
\end{equation}
equivalent to one-dimensional particle motion where $V_{\rm eff}$ is the effective potential:
\begin{equation}
    V_{\rm eff}(r)=\frac{\mathrm{L}^2}{r^2}\,f(r)=\frac{\mathrm{L}^2}{r^2}\left(1-8\pi K-\frac{2 M}{r}+\frac{4\pi K \lambda}{r^2}\right).\label{bb4}
\end{equation}

\begin{figure}[ht!]
    \centering
    \includegraphics[width=0.8\linewidth]{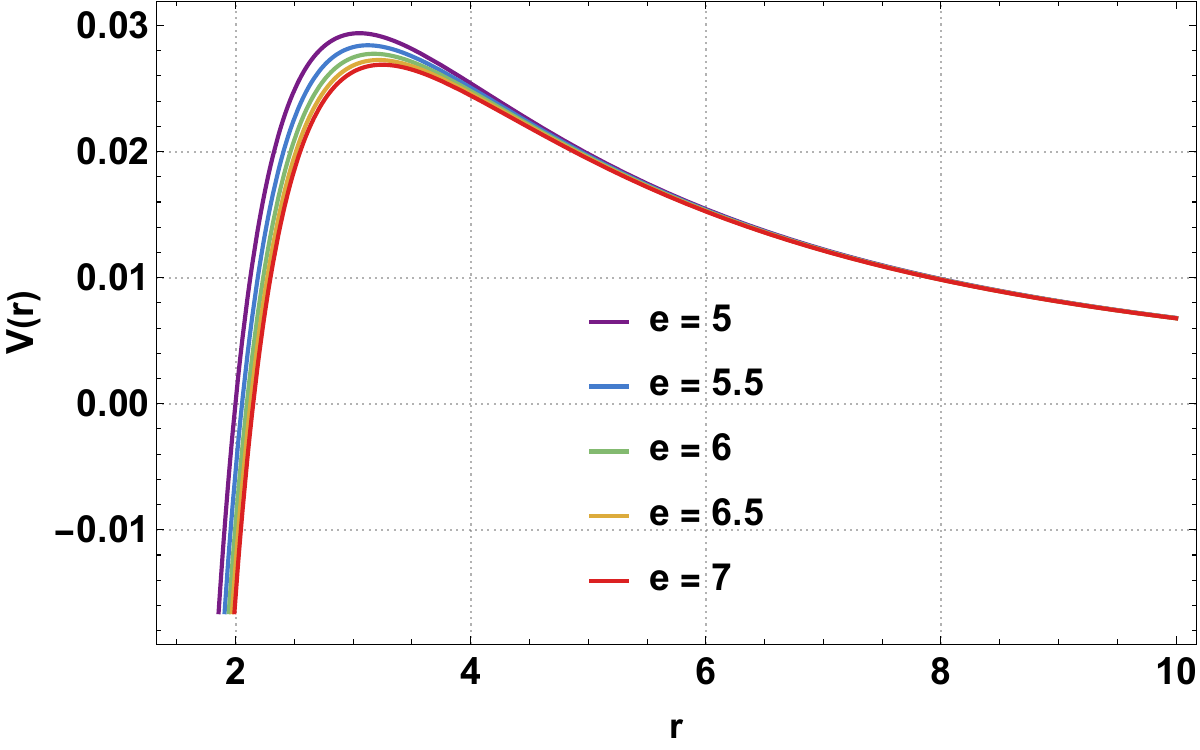}\\
     (i) $F_{\pi}=0.141$ \\
     \hfill\\
    \includegraphics[width=0.8\linewidth]{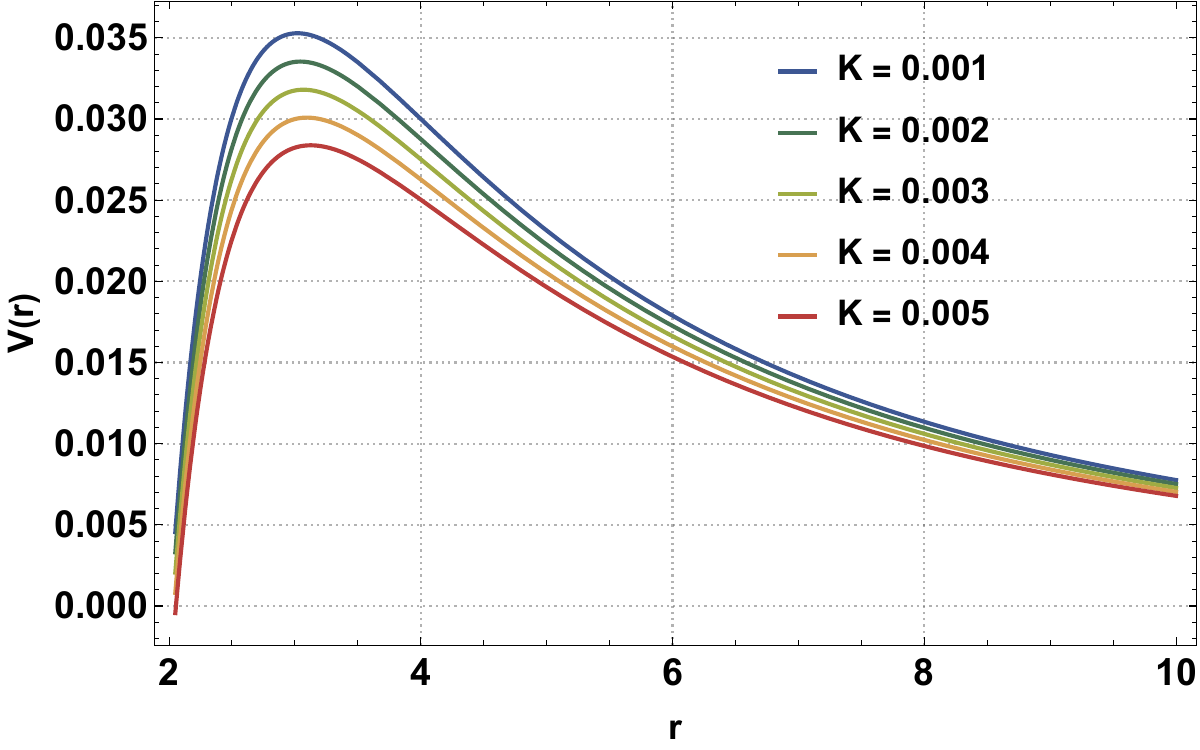}\\
   (ii) $e=5.5$
    \caption{Effective potential versus radial distance $r$ for varying coupling parameter $e$ (upper panel) and Skyrme parameter $K$ (lower panel). Here $M=1=\mathrm{L}$.}
    \label{fig:null}
\end{figure}

This effective potential governs optical features including the photon sphere, BH shadow, photon trajectories, and effective force on massless particles.

Figure~\ref{fig:null} displays the effective potential as a function of radial coordinate $r$ for various coupling parameter $e$ and Skyrme parameter $K$ values. Both panels show the effective potential peak decreasing as either $e$ or $K$ increases, indicating a weakening gravitational potential barrier. This reduced barrier affects circular null orbit location and stability, requires lower photon energy for escape, and consequently modifies observable features including shadow size and shape and the critical impact parameter separating captured from scattered photon paths.

\begin{center}
    {A.\, \bf Circular Orbits: Photon Sphere}
\end{center}

The photon sphere is where photons execute unstable circular orbits due to strong gravitational effects. Parameters $K$ and $\lambda$ modify the spacetime curvature and consequently alter the photon sphere.

Circular null orbits require $\dot{r}=0$ and $\ddot{r}=0$, which via Eq.~(\ref{bb3}) implies
\begin{equation}
    \mathrm{E}^2=V_{\rm eff}\quad \mbox{and} \quad \partial_r V_{\rm eff}(r)=0.\label{cc1}
\end{equation}

The first condition yields the critical impact parameter:
\begin{equation}
    \beta_c=\frac{\mathrm{L}}{\mathrm{E}}\Big{|}_{r=r_{\rm ph}}=\frac{r^2_{\rm ph}}{\sqrt{(1-8\pi K) r^2_{\rm ph}-2 M r_{\rm ph}+4\pi K \lambda}},\label{cc2}
\end{equation}
where $r_{\rm ph}$ is the photon sphere radius obtained from the second condition in Eq.~(\ref{cc1}):
\begin{align}
    r_{\rm ph}=\frac{3 M+\sqrt{9 M^2-32\pi K \lambda (1-8\pi K)}}{2 (1-8\pi K)}.\label{cc3}
\end{align}
Photon sphere existence requires
\begin{equation}
    32\pi K \lambda (1-8\pi K) \leq 9 M^2.\label{cc4}
\end{equation}

\begin{figure}[ht!]
    \centering
    \includegraphics[width=0.8\linewidth]{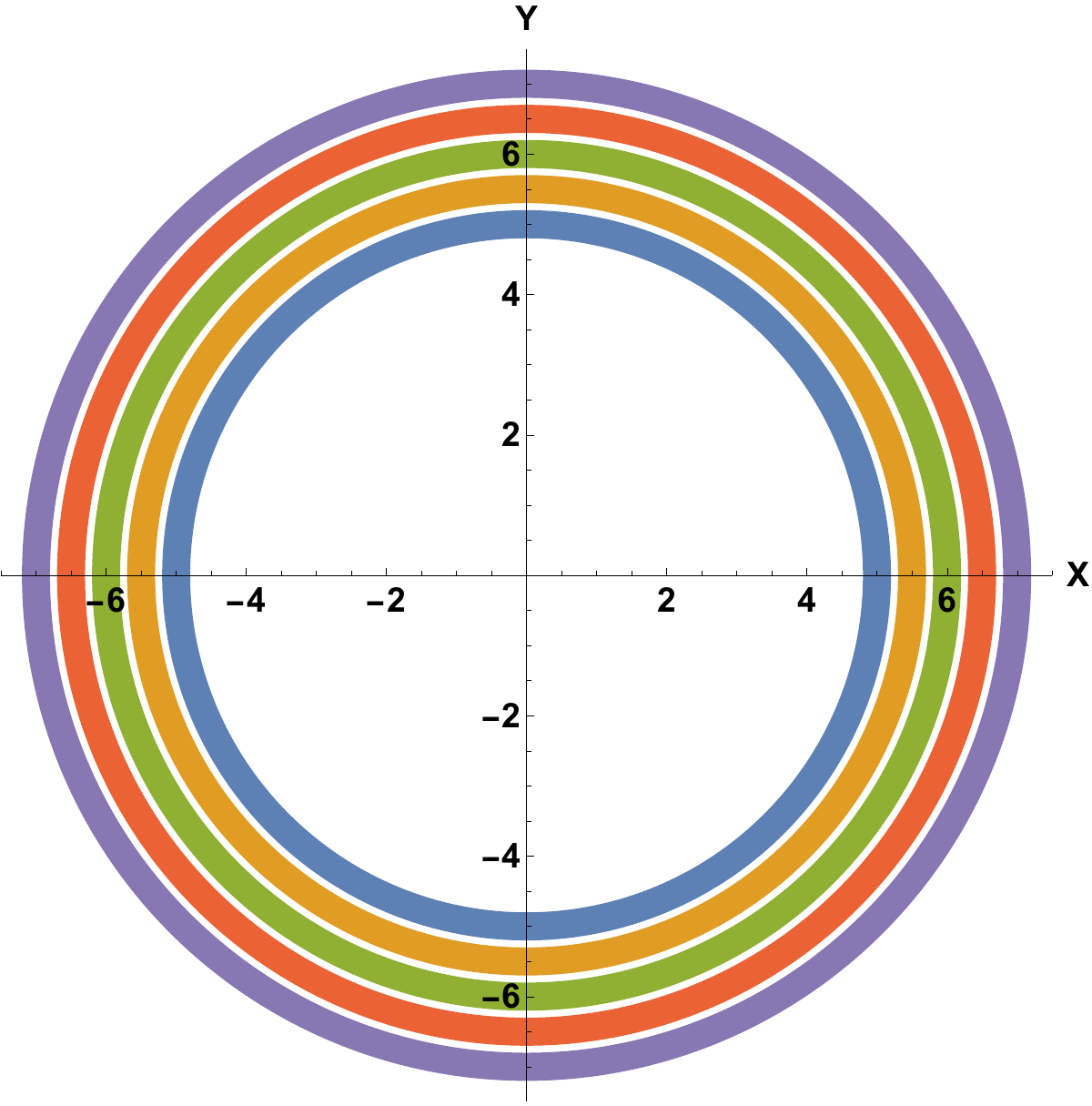}
    \caption{Photon spheres for varying $e$ with $M = 1$ and $F_{\pi} = 0.141$. From inner to outer rings: $e = 5$ to $7$ in steps of $0.5$.}
    \label{fig:ring}
\end{figure}

Figure~\ref{fig:ring} illustrates the photon sphere variation with coupling parameter $e$. The rings expand outward as $e$ increases from 5 to 7, reflecting the photon sphere radius dependence on the quartic Skyrme parameter $\lambda = 4/(e^2 F_\pi^2)$.

\begin{center}
    {B.\, \bf Shadow Radius}
\end{center}

The BH shadow forms from photon capture within the photon sphere, with size and shape depending on geometric parameters $K$ and $\lambda$ and BH mass $M$.

At large distances, the lapse function behaves as
\begin{equation}
    \lim_{r \to \infty} f(r) =1-8\pi K \neq 1,\label{cc5}
\end{equation}
indicating asymptotically local but not global flatness. The Skyrme coupling constant $K$ produces a solid angle deficit analogous to the global monopole case~\cite{Barriola1989}.

Following~\cite{Volker2022}, the shadow angular size for a static observer at position $r_O$ is
\begin{equation}
    \sin^2 \vartheta_{\rm sh}=\frac{h(r_{\rm ph})}{h(r_O)},\quad h(r)=\frac{r}{\sqrt{f(r)}}.\label{cc6}
\end{equation}
The shadow radius ($R_{\rm sh} \simeq r_O \vartheta_{\rm sh}$) is
\begin{align}
    R_{\rm sh}=r_{\rm ph}\sqrt{\frac{f(r_O)}{f(r_{\rm ph})}}=r_{\rm ph} \sqrt{\frac{1-8\pi K-\frac{2 M}{r_O}+\frac{4\pi K \lambda}{r^2_O}}{1-8\pi K-\frac{2 M}{r_{\rm ph}}+\frac{4\pi K \lambda}{r^2_{\rm ph}}}},\label{cc7}
\end{align}
where $r_{\rm ph}$ is given by Eq.~(\ref{cc3}).

\begin{figure}[ht!]
    \centering
    \includegraphics[width=0.8\linewidth]{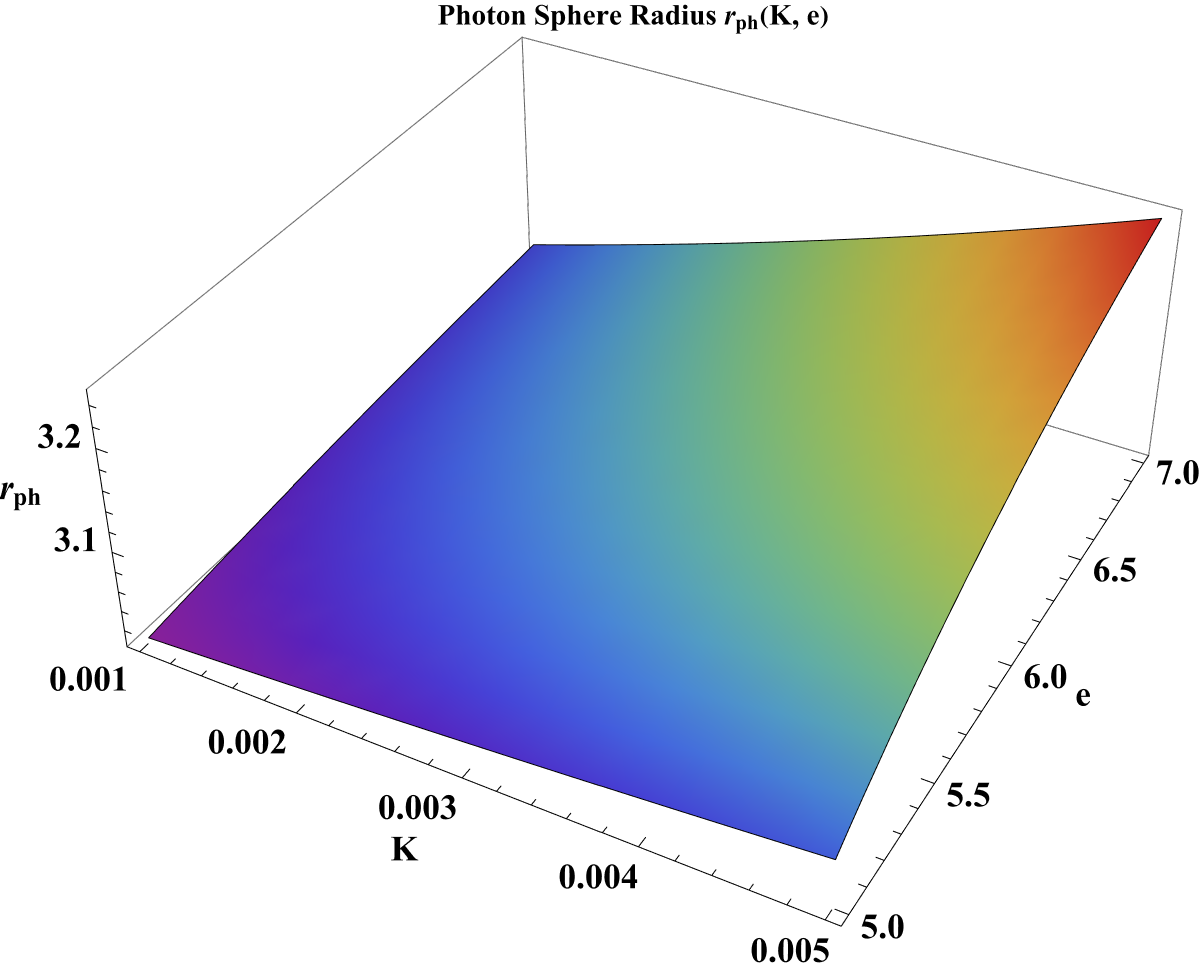}\\
    \hfill\\
    \includegraphics[width=0.8\linewidth]{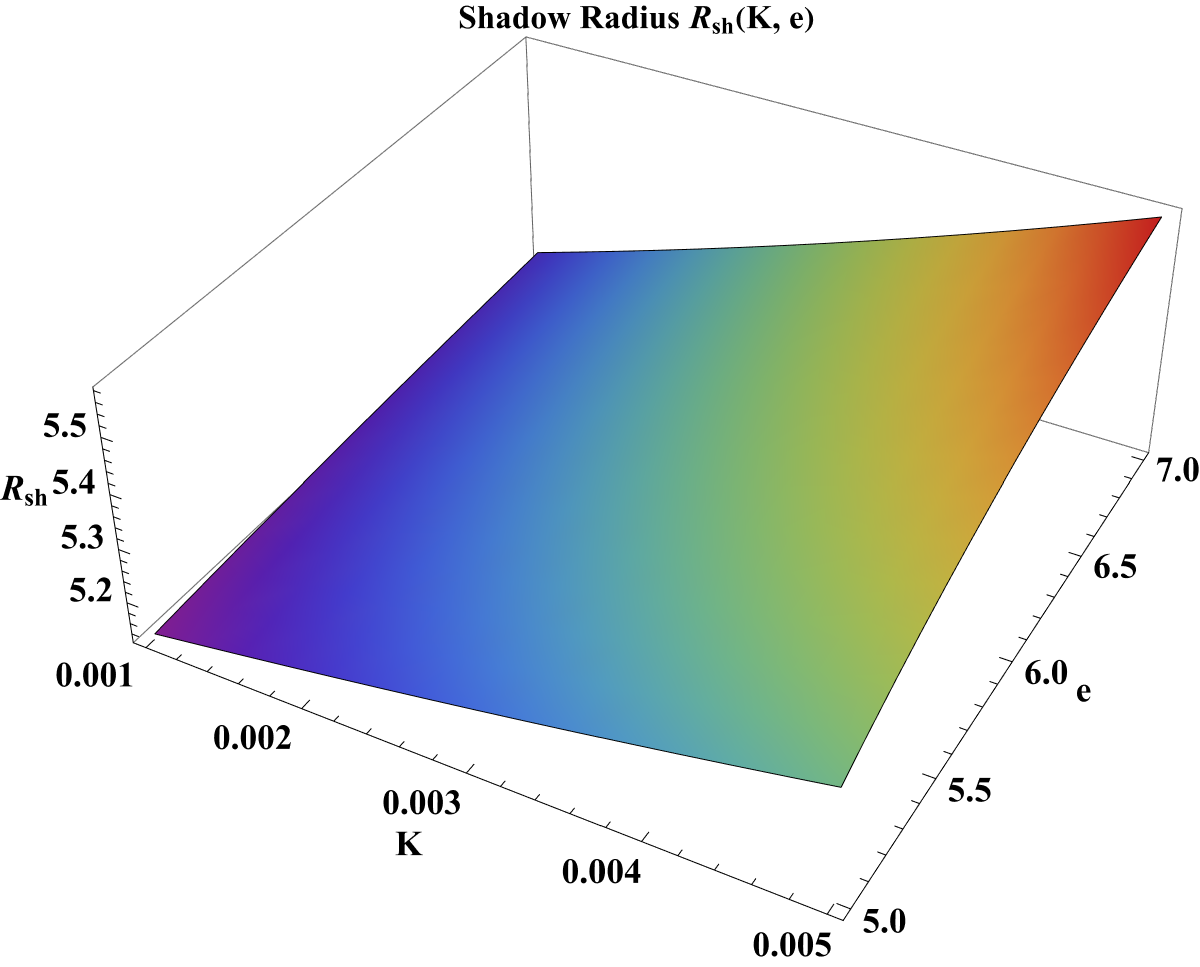}
    \caption{Three-dimensional surfaces showing photon sphere radius $r_{\rm ph}$ (upper) and shadow radius $R_{\rm sh}$ (lower) as functions of $K$ and $e$.}
    \label{fig:photon-shadow}
\end{figure}

For a distant static observer, the shadow radius simplifies to
\begin{equation}
    R_{\rm sh}=r_{\rm ph} \sqrt{\frac{1- 8\pi K}{1-8\pi K-\frac{2 M}{r_{\rm ph}}+\frac{4\pi K \lambda}{r^2_{\rm ph}}}}.\label{cc8}
\end{equation}

Expressions~(\ref{cc3}) and~(\ref{cc8}) show that parameters $K$ and $\lambda$ modify the photon sphere and shadow radii relative to the standard BH. Setting $\lambda=0$ yields
\begin{equation}
r_{\rm ph}=\frac{3 M}{1-8\pi K}\quad, \quad R_{\rm sh}=\sqrt{3}\, r_{\rm ph}=\frac{3\sqrt{3} M}{1-8\pi K},\label{cc9}
\end{equation}
recovering the global monopole-like spacetime results.

\begin{figure*}[t!]
\centering
\includegraphics[width=0.9\linewidth]{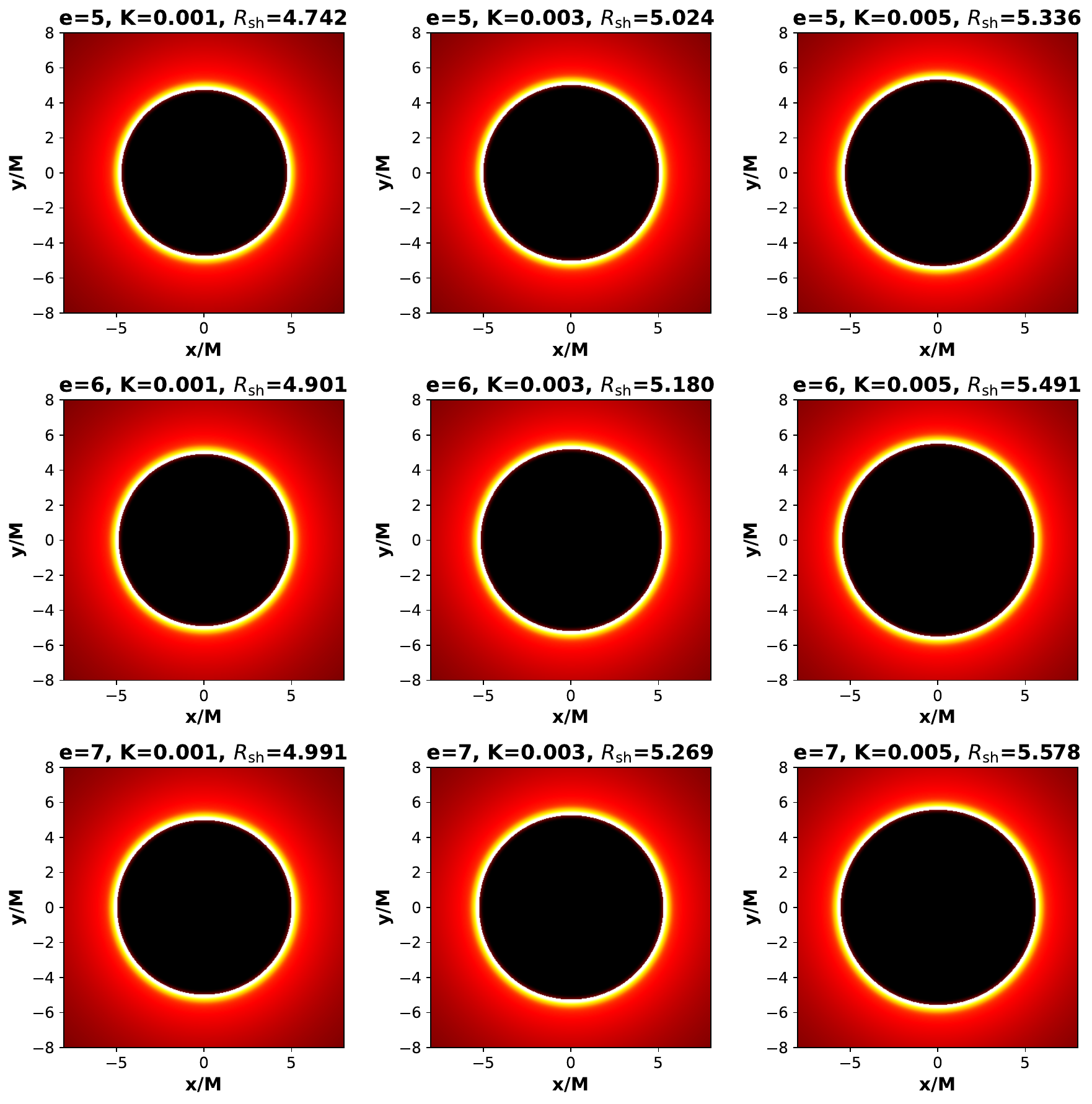}
\caption{BH shadow intensity maps for $e=5,6,7$ (rows) and $K=0.001,0.003,0.005$ (columns). Shadow radii $R_{\rm sh}/M$ are computed from Eqs.~(\ref{cc3}) and~(\ref{cc7}) with $r_O=50$.}
\label{fig:shadow-grid}
\end{figure*}

Figure~\ref{fig:photon-shadow} presents three-dimensional surfaces of the photon sphere and shadow radii as functions of $K$ and $e$. Both radii increase with $K$ and decrease with $e$, consistent with Eqs.~(\ref{cc3}) and~(\ref{cc8}). Numerical values for the horizon, photon sphere and shadow radii are collected in Table~\ref{tab:shadow}, which scans both the Skyrme coupling $K$ and the pion coupling $e$ over the phenomenologically relevant ranges. All entries satisfy the constraint $K\lambda=1/e^{2}$ inherited from the underlying theory, and the nine combinations marked with an asterisk ($^*$) correspond to the intensity maps displayed in Fig.~\ref{fig:shadow-grid}.

\begin{table}[ht!]
\centering
\renewcommand{\arraystretch}{1.15}
\setlength{\tabcolsep}{4pt}
\begin{tabular}{|cc|cccc|}
\hline
$K$ & $e$ & $\lambda$ & $r_h/M$ & $r_{\rm ph}/M$ & $R_{\rm sh}/M$ \\
\hline
\multicolumn{6}{|c|}{Varying $K$ at fixed $e$} \\
\hline
0.001$^*$ & 5 & 40.0000 & 1.75832 & 2.69465 & 4.74249 \\
0.002 & 5 & 20.0000 & 1.81411 & 2.77770 & 4.87957 \\
0.003$^*$ & 5 & 13.3333 & 1.87281 & 2.86515 & 5.02381 \\
0.004 & 5 & 10.0000 & 1.93468 & 2.95737 & 5.17580 \\
0.005$^*$ & 5 & 8.0000 & 2.00000 & 3.05478 & 5.33620 \\
\hline
0.001$^*$ & 6 & 27.7778 & 1.85894 & 2.82373 & 4.90112 \\
0.002 & 6 & 13.8889 & 1.91380 & 2.90581 & 5.03711 \\
0.003$^*$ & 6 & 9.2593 & 1.97161 & 2.99231 & 5.18026 \\
0.004 & 6 & 6.9444 & 2.03261 & 3.08359 & 5.33119 \\
0.005$^*$ & 6 & 5.5556 & 2.09707 & 3.18009 & 5.49053 \\
\hline
0.001$^*$ & 7 & 20.4082 & 1.91413 & 2.89564 & 4.99061 \\
0.002 & 7 & 10.2041 & 1.96869 & 2.97739 & 5.12619 \\
0.003$^*$ & 7 & 6.8027 & 2.02620 & 3.06356 & 5.26894 \\
0.004 & 7 & 5.1020 & 2.08691 & 3.15453 & 5.41946 \\
0.005$^*$ & 7 & 4.0816 & 2.15109 & 3.25071 & 5.57840 \\
\hline
\multicolumn{6}{|c|}{Varying $e$ at physical $K=F_{\pi}^2/4\simeq 4.97\times 10^{-3}$} \\
\hline
0.00497 & 5.0 & 8.0479 & 1.99800 & 3.05180 & 5.33130 \\
0.00497 & 5.5 & 6.6511 & 2.05442 & 3.12436 & 5.42039 \\
0.00497 & 6.0 & 5.5888 & 2.09510 & 3.17714 & 5.48566 \\
0.00497 & 6.5 & 4.7621 & 2.12559 & 3.21693 & 5.53510 \\
0.00497 & 7.0 & 4.1061 & 2.14913 & 3.24777 & 5.57354 \\
\hline
\end{tabular}
\caption{Horizon, photon sphere and shadow radii of the Skyrmion BH with $M=1$ and $r_O=50$. Upper panel: $K$ varies at fixed $e\in\{5,6,7\}$. Lower panel: $e$ varies at the physical value $K=F_{\pi}^2/4$. Entries marked $^*$ correspond to Fig.~\ref{fig:shadow-grid}. The Schwarzschild reference is $R_{\rm sh}/M=3\sqrt{3}\simeq 5.196$.}
\label{tab:shadow}
\end{table}
 
Figure~\ref{fig:shadow-grid} displays the optical appearance of the Skyrmion BH across a $3\times 3$ parameter grid using the inferno colourscheme. Each panel shows a dark silhouette surrounded by a bright photon ring, with the shadow radius $R_{\rm sh}/M$ printed above. As $K$ increases from $0.001$ to $0.005$ along any row, the shadow grows noticeably larger: $R_{\rm sh}/M$ rises from $4.74$ to $5.34$ for $e=5$, from $4.90$ to $5.49$ for $e=6$, and from $4.99$ to $5.58$ for $e=7$. Movement along columns reveals a weaker $e$ dependence, confirming that the Skyrme coupling $K$ dominates the shadow response. The smallest shadow appears at $(e,K)=(5,0.001)$ and the largest at $(e,K)=(7,0.005)$.

Table~\ref{tab:shadow} shows that the shadow radius grows from $5.331$ at $e=5$ to $5.573$ at $e=7$ when $K$ is fixed at its physical value, all exceeding the Schwarzschild benchmark $3\sqrt{3}\simeq 5.196$. Two limiting checks verify the expressions: sending $K\to 0$ with $\lambda$ finite recovers $r_{\rm ph}/M\to 3$ and $R_{\rm sh}/M\to 3\sqrt{3}$, while holding $K\lambda=1/36$ fixed and taking $K\to 0$ yields a Reissner--Nordstr\"om-like geometry with $r_{\rm ph}/M\simeq 2.746$ and $R_{\rm sh}/M\simeq 4.870$.

\begin{center}
    {C.\, \bf Effective Radial Force}
\end{center}

The effective radial force on photons determines whether they are captured or escape to infinity, given by the negative effective potential gradient:
\begin{equation}
    \mathrm{F}_{\rm ph}=-\frac{1}{2} \partial_r V_{\rm eff}(r)=\frac{\mathrm{L}^2}{r^3}\left(1-8\pi K-\frac{3 M}{r}+\frac{16\pi K \lambda}{r^2}\right).\label{dd1}
\end{equation}

Parameters $K$ and $\lambda$ influence this force, as does the conserved angular momentum $\mathrm{L}$.

\begin{figure*}[ht!]
    \centering
    \includegraphics[width=0.48\linewidth]{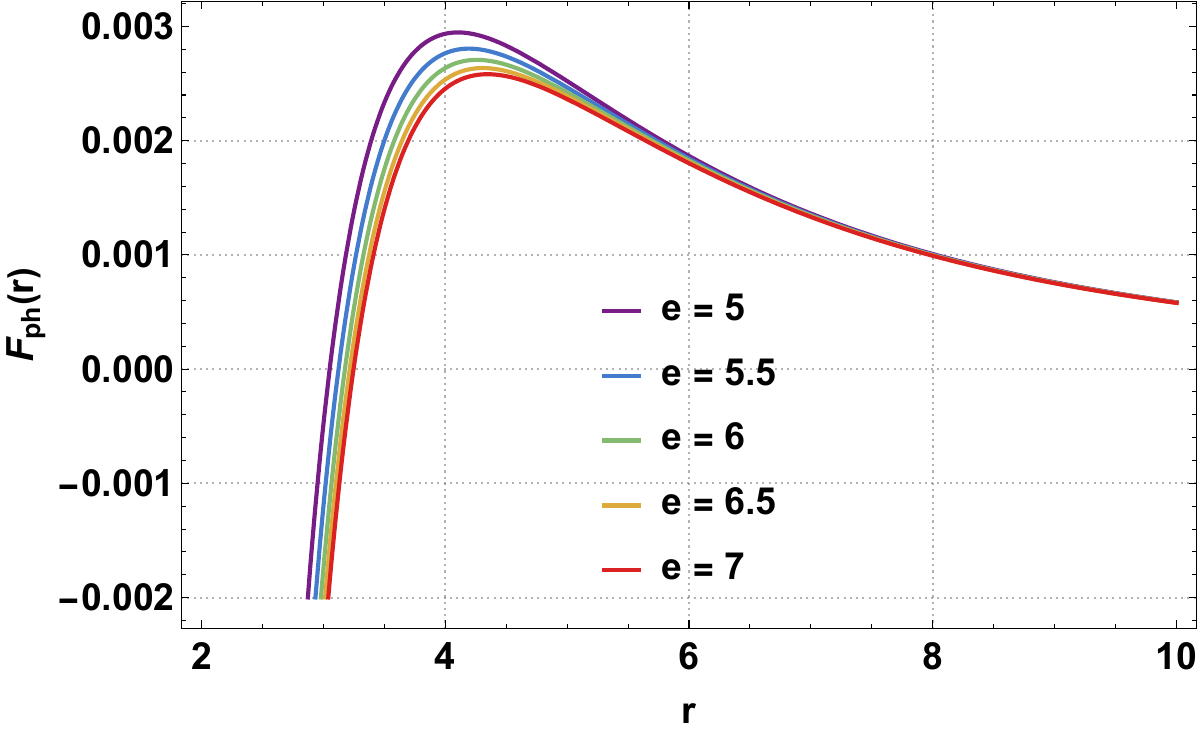}
    \hfill
    \includegraphics[width=0.48\linewidth]{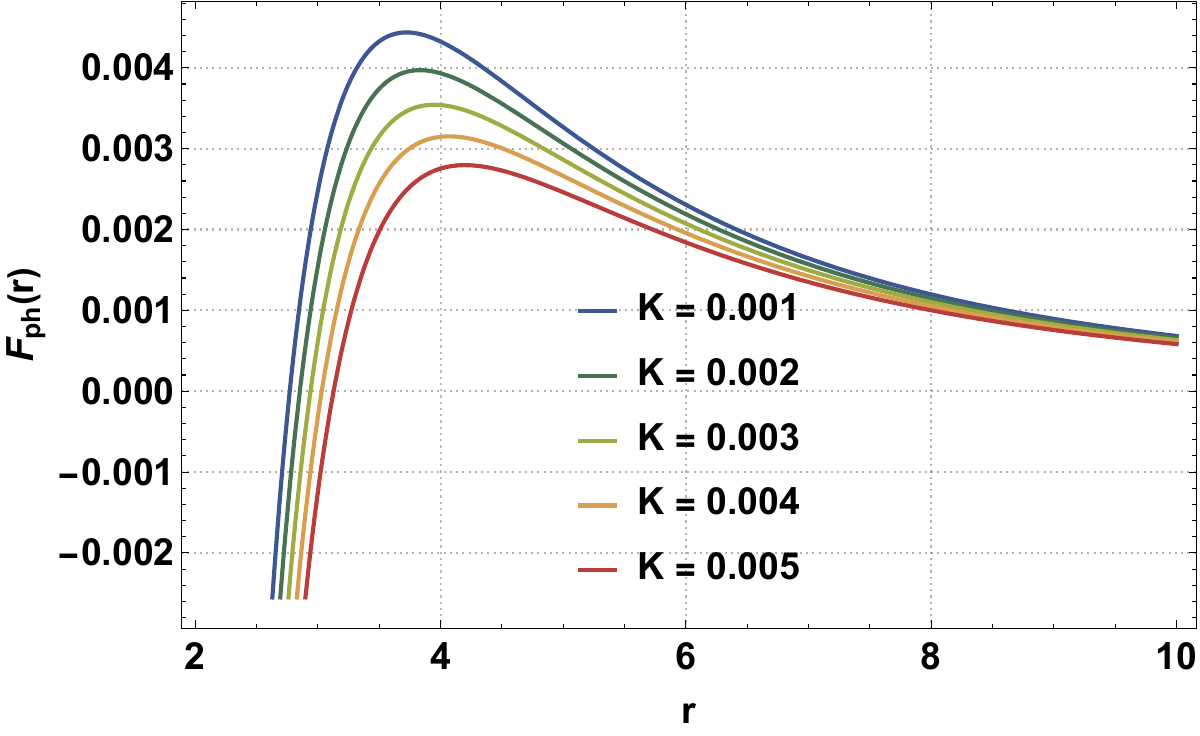}\\[1mm]
    \hspace{0.25\linewidth}(i) $F_{\pi}=0.141$
    \hfill
    (ii) $e=5.5$\hspace{0.25\linewidth}
    \caption{Effective radial force on photons versus radial distance $r$ for varying $e$ (left) and $K$ (right). Here $M=1=\mathrm{L}$.}
    \label{fig:force}
\end{figure*}

Figure~\ref{fig:force} shows the effective radial force versus $r$ for different $e$ and $K$ values. Both panels exhibit decreasing force peaks as $e$ or $K$ increases, reflecting reduced gravitational barriers that enhance photon escape probability and modify trajectory characteristics.

\begin{figure*}[ht!]
    \centering
    \includegraphics[width=0.3\linewidth]{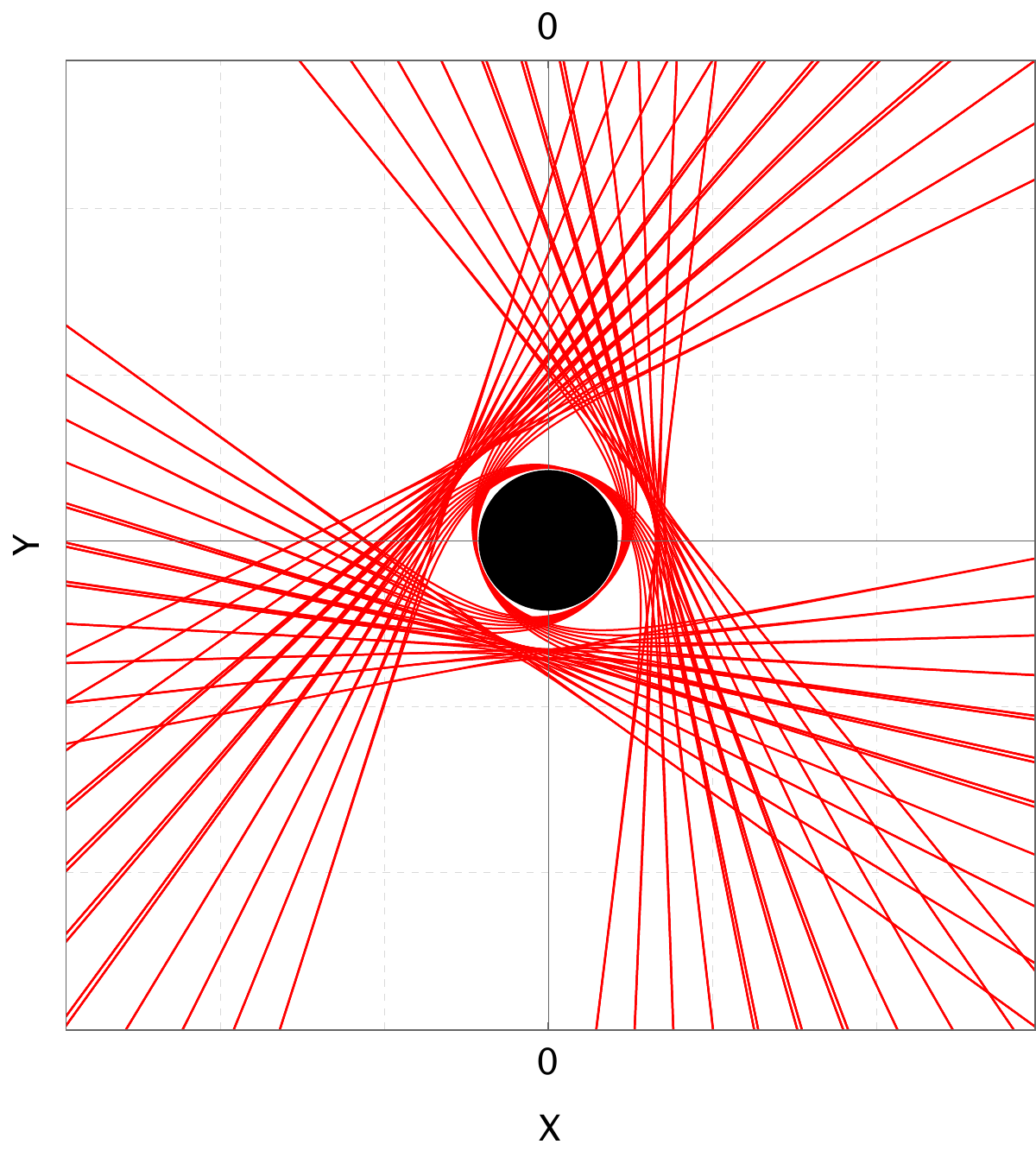}\quad
    \includegraphics[width=0.3\linewidth]{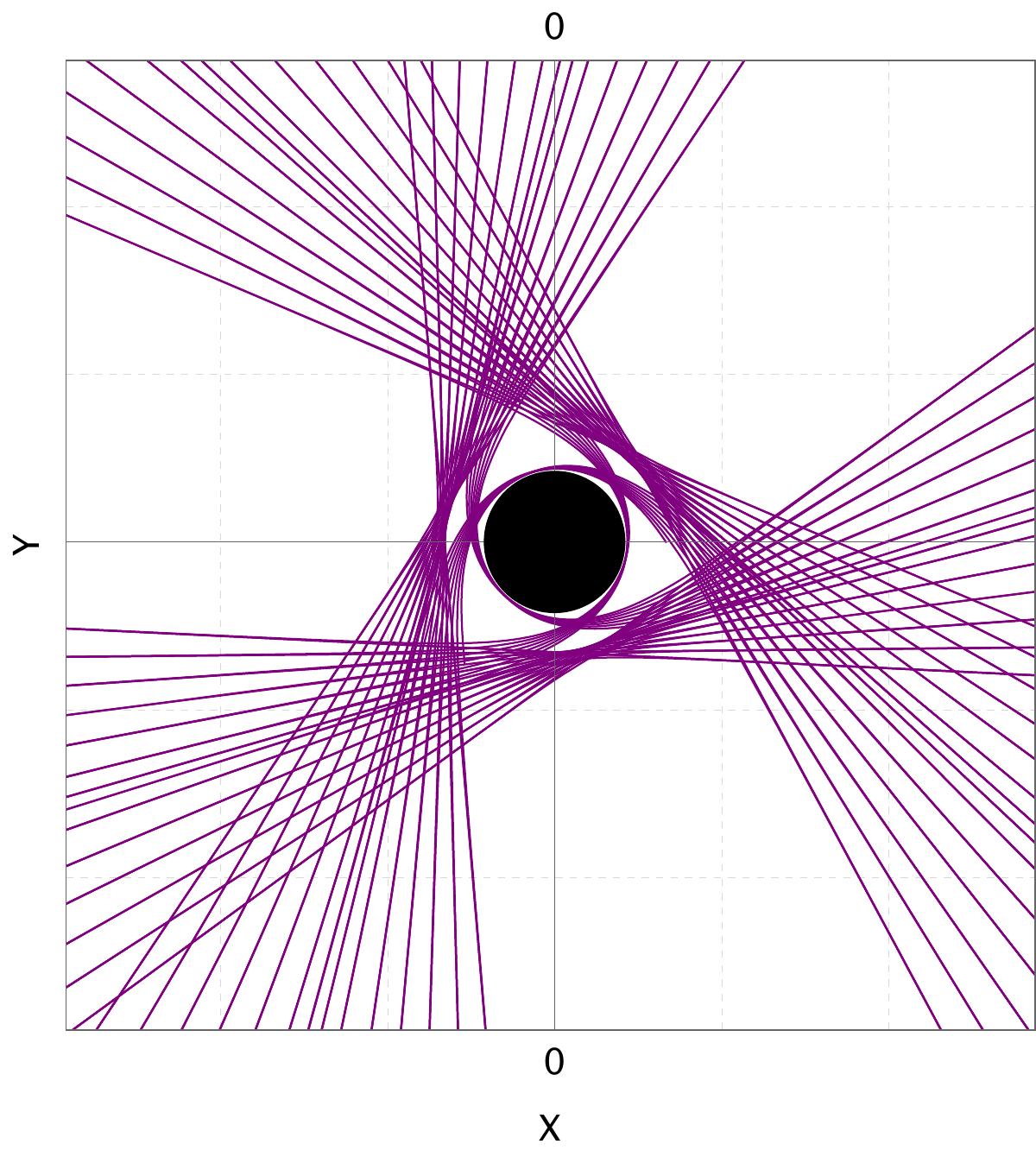}\quad
    \includegraphics[width=0.3\linewidth]{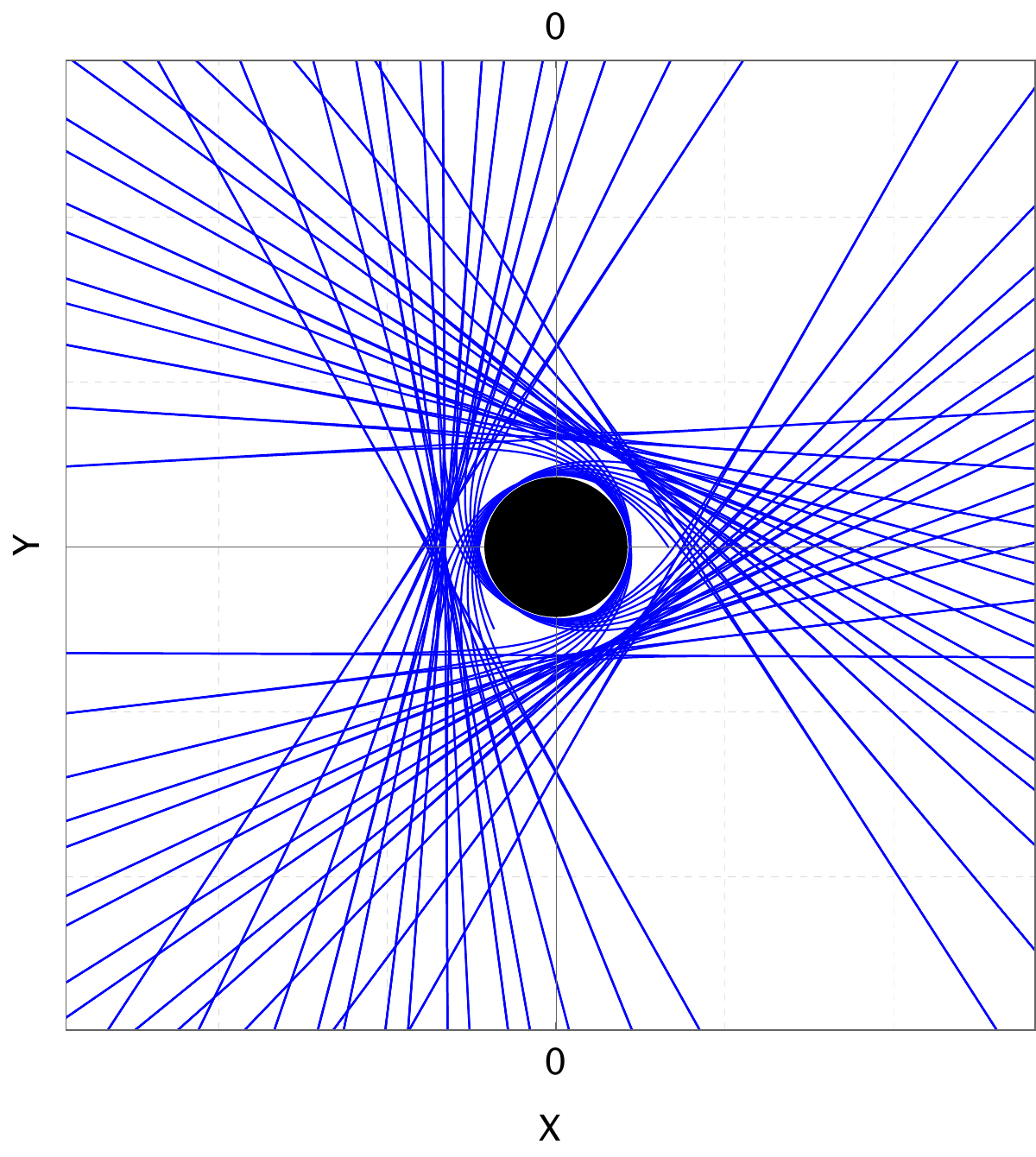}\\
    (i) $e=5$ \hspace{4cm} (ii) $e=5.2$ \hspace{4cm} (iii) $e=5.3$ \\
    \hfill\\
    \centering
    \includegraphics[width=0.3\linewidth]{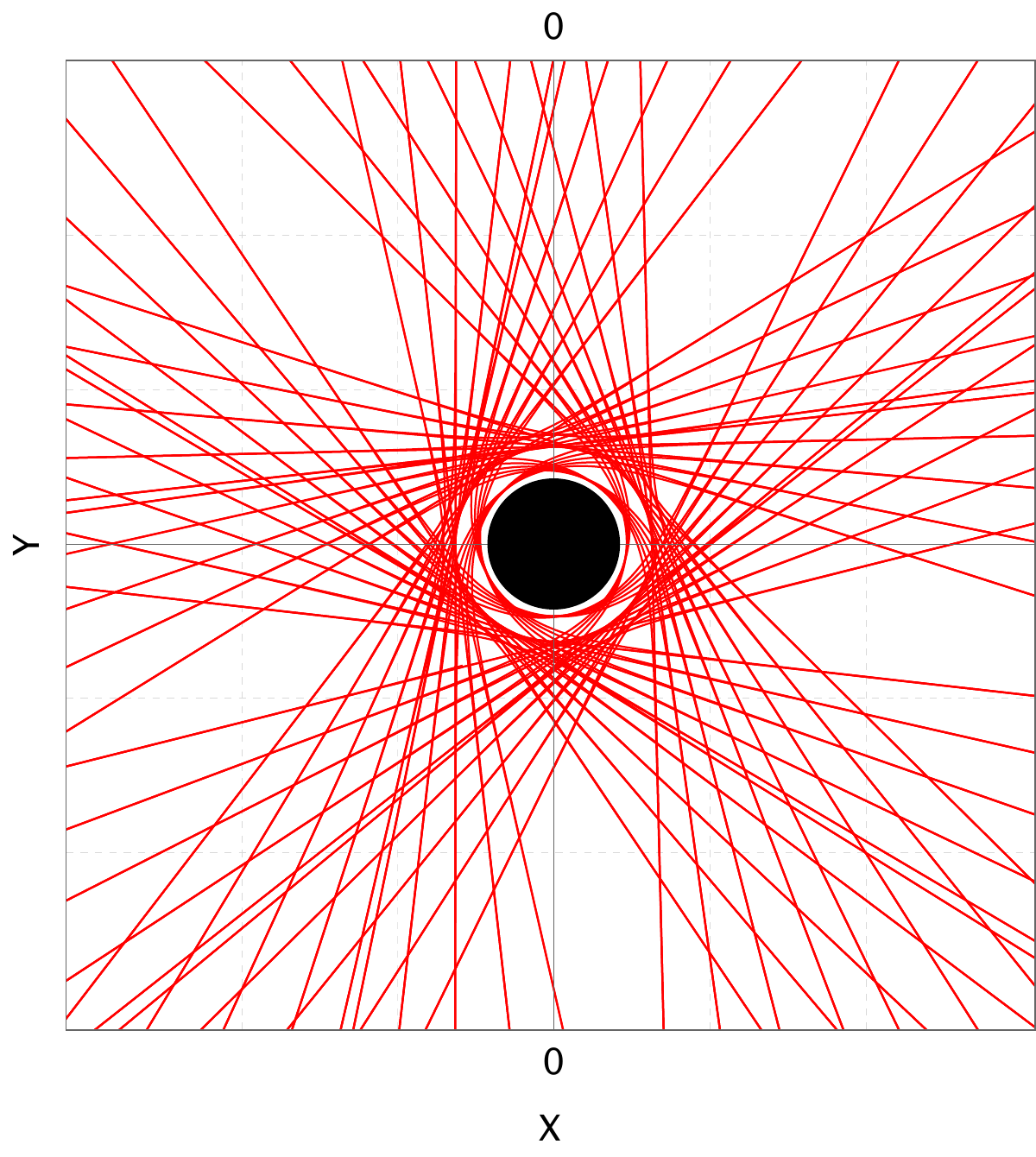}\quad
    \includegraphics[width=0.3\linewidth]{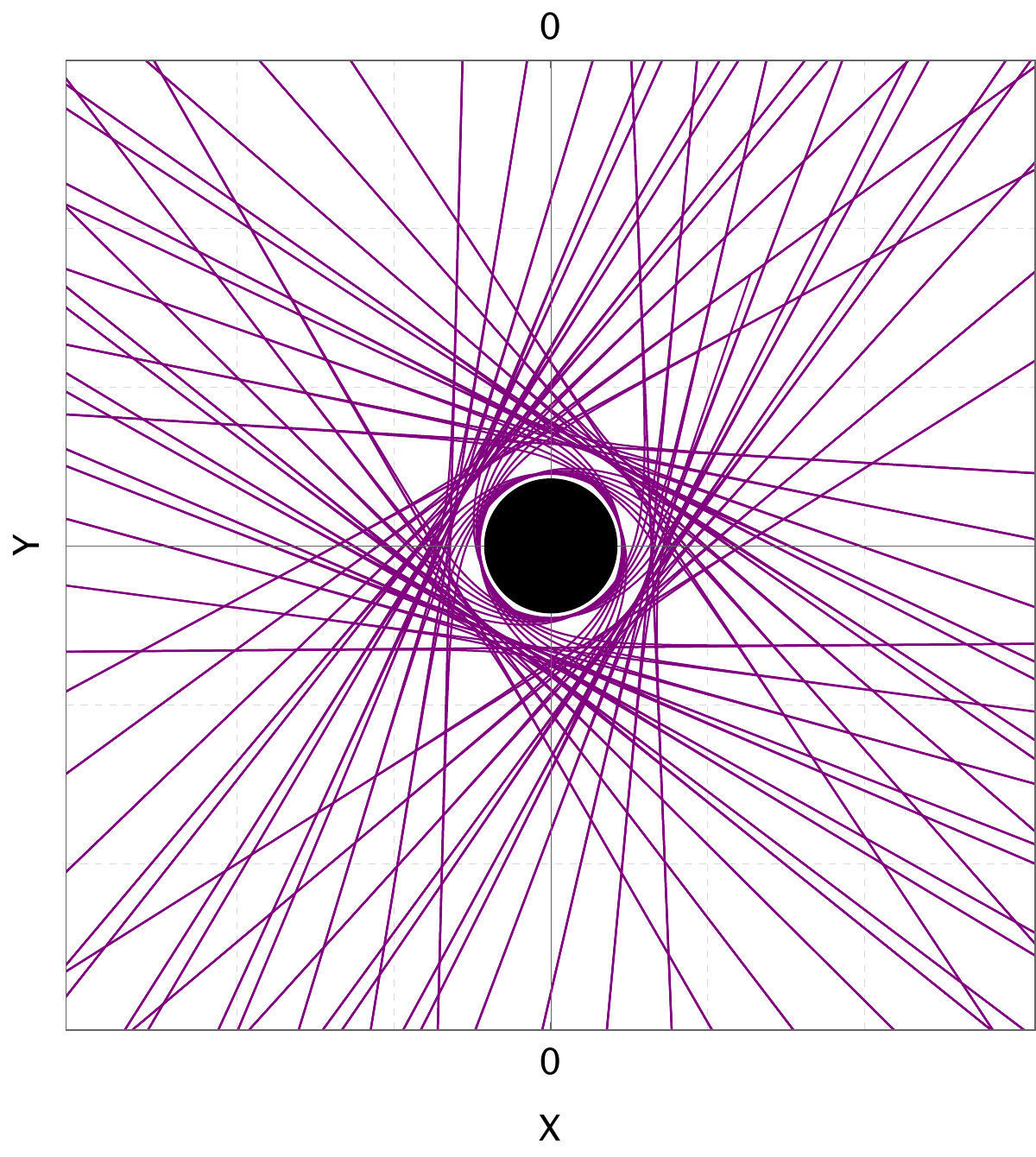}\quad
    \includegraphics[width=0.3\linewidth]{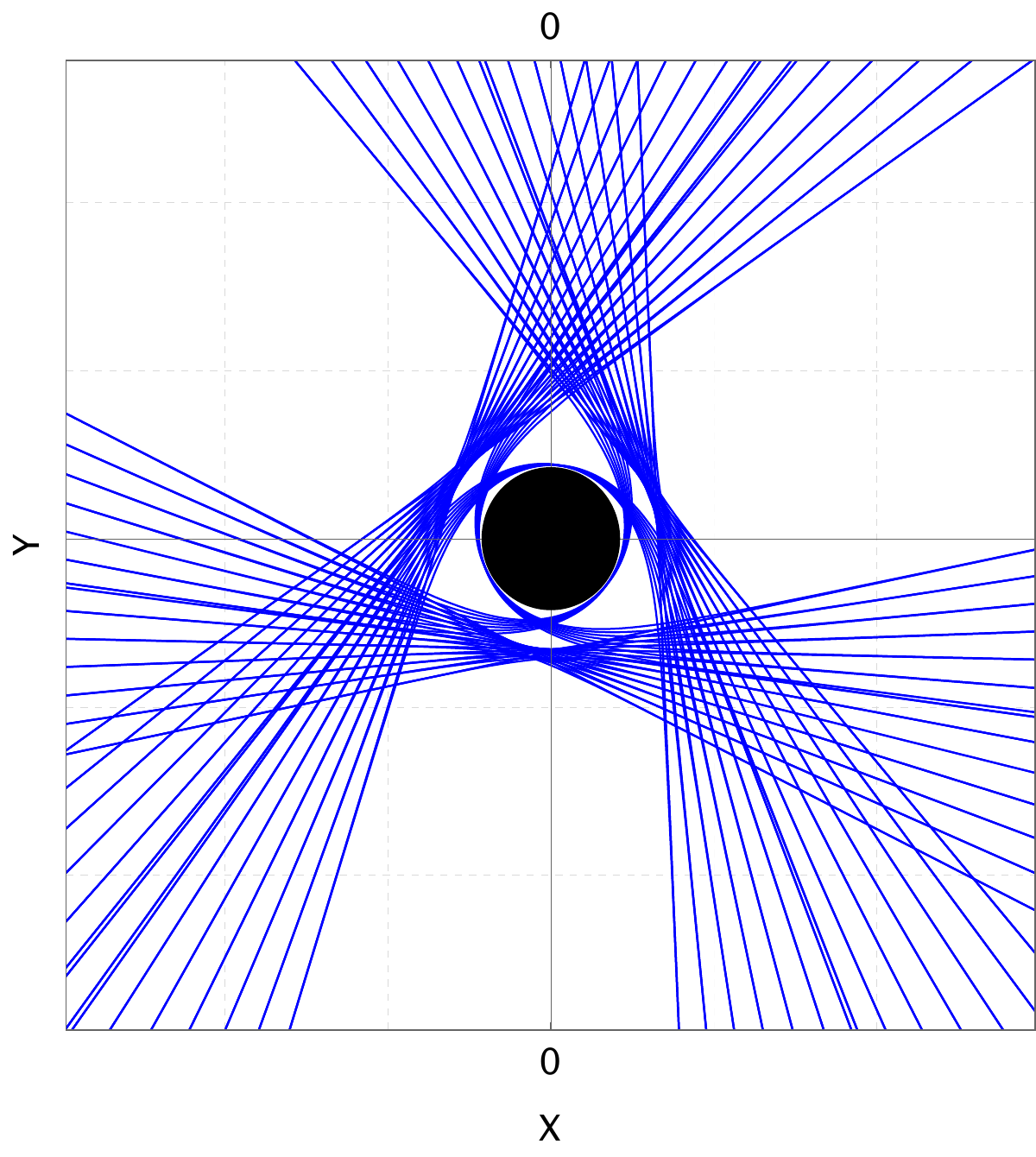}\\
    (i) $K=0.001$ \hspace{3cm}  (ii) $K=0.003$ \hspace{3cm} (iii) $K=0.005$ 
    \caption{Photon trajectories from Eq.~(\ref{ee3}) for different $e$ values (upper row, $F_\pi=0.141$) and $K$ values (lower row, $e=5$). Here $M=1$.}
    \label{fig:parametric}
\end{figure*}

\begin{center}
    {D.\, \bf Photon Trajectories}
\end{center}

We now examine photon trajectories in the Skyrme BH gravitational field and how coupling constants $K$ and $\lambda$ modify them.

Using Eqs.~(\ref{bb2}) and~(\ref{bb3}) with $\eta^2=8\pi K$, $r_s=2 M$, and $Q^2=4\pi K \lambda$, the orbit equation becomes
\begin{equation}
    \left(\frac{1}{r^2}\frac{dr}{d\phi}\right)^2+\frac{1-\eta^2}{r^2}=\frac{1}{\beta^2}+\frac{r_s}{r^3}-\frac{Q^2}{r^4},\label{ee1}
\end{equation}
where $\beta=\mathrm{L}/\mathrm{E}$ is the photon impact parameter.

Transforming via $u(\phi)=1/r(\phi)$ yields
\begin{equation}
    \left(\frac{du}{d\phi}\right)^2+(1-\eta^2) u^2=\frac{1}{\beta^2}+r_s u^3-Q^2 u^4.\label{ee2}
\end{equation}
Differentiation gives
\begin{equation}
\frac{d^2u}{d\phi^2}+(1-\eta^2) u=\frac{3 r_s}{2} u^2-2 Q^2 u^3.\label{ee3}
\end{equation}

This second-order equation governs Skyrme BH photon trajectories, with $K$ and $\lambda$ significantly influencing trajectory behavior.

Figure~\ref{fig:parametric} depicts photon trajectories for various $e$ and $K$ values with initial conditions $u(0) = 0.15$, $u'(0) = 0.1$, and $M = 1$. The upper row fixes $F_{\pi} = 0.141$ while varying $e$; the lower row fixes $e = 5$ while varying $K$. The trajectories show systematic changes in deflection patterns as parameters vary.

A perturbative weak-field solution $u \simeq u_0+\epsilon u_1$ gives the zeroth-order result:
\begin{equation}
u_0(\phi) = \frac{1}{\beta} \cos(\omega \phi), \quad \omega = \sqrt{1-\eta^2},\label{ee4}
\end{equation}
and first-order correction:
\begin{align}
u_1(\phi) &= \frac{3 r_s}{4 \omega^2 \beta^2} - \frac{r_s}{4 \omega^2 \beta^2} \cos(2 \omega \phi) - \frac{3 Q^2}{4 \omega \beta^3} \phi \sin(\omega \phi)\nonumber\\
&- \frac{Q^2}{16 \omega^2 \beta^3} \cos(3 \omega \phi),\label{ee5}
\end{align}
where the secular term $-\frac{3 Q^2}{4 \omega \beta^3} \phi \sin(\omega \phi)$ produces perihelion precession.

The complete perturbative solution of Eq.~(\ref{ee3}) is
\begin{widetext}
\begin{align}
\frac{1}{r(\phi)}=u(\phi) &= \frac{1}{\beta} \cos \left(\sqrt{1-8\pi K}\, \phi\right) + \epsilon \Bigg[
\frac{3 r_s}{4 (1-8\pi K) \beta^2} - \frac{r_s}{4 (1-8\pi K) \beta^2} \cos \left(2 \sqrt{1-8\pi K}\, \phi\right)\nonumber\\
&- \frac{3\pi K \lambda\, \phi}{\sqrt{1-8\pi K}\, \beta^3}\, \sin \left(\sqrt{1-8\pi K}\, \phi\right)- \frac{\pi K \lambda}{4 (1-8\pi K)\beta^3} \cos \left(3 \sqrt{1-8\pi K}\, \phi\right)
\Bigg].\label{ee6}
\end{align}
\end{widetext}

Using first-order perturbation theory, the explicit approximate expression for the deflection angle is (where the impact parameter $\beta \gg r_s$)
\begin{align}
\hat{\alpha} &\simeq \pi \left( \frac{1}{\sqrt{1-8\pi K}} - 1 \right) + \frac{4 M}{\beta} - \frac{3\pi^2 K \lambda}{\beta^2}\nonumber\\
& \simeq 4\pi^2 K+ \frac{4 M}{\beta} - \frac{3\pi^2 K \lambda}{\beta^2}.\label{ee7}    
\end{align}

The first term $4\pi^2 K$ arises from the Skyrme parameter and represents a constant angular offset independent of impact parameter. The second term $4M/\beta$ recovers the standard Schwarzschild deflection, while the third term $-3\pi^2 K\lambda/\beta^2$ introduces a quartic Skyrme correction that becomes significant at small impact parameters.

Setting $\lambda=0$ gives
\begin{equation}
\hat{\alpha} \simeq 4\pi^2 K+ \frac{4 M}{\beta},\label{ee8}    
\end{equation}
analogous to a Schwarzschild BH that absorbed a global monopole. This limiting case connects directly to the shadow analysis in Sec.~II\,B, where Eq.~(\ref{cc9}) showed that the photon sphere and shadow radii also reduce to global monopole-like expressions when $\lambda \to 0$.

\begin{figure}[ht!]
\centering
\includegraphics[width=0.8\linewidth]{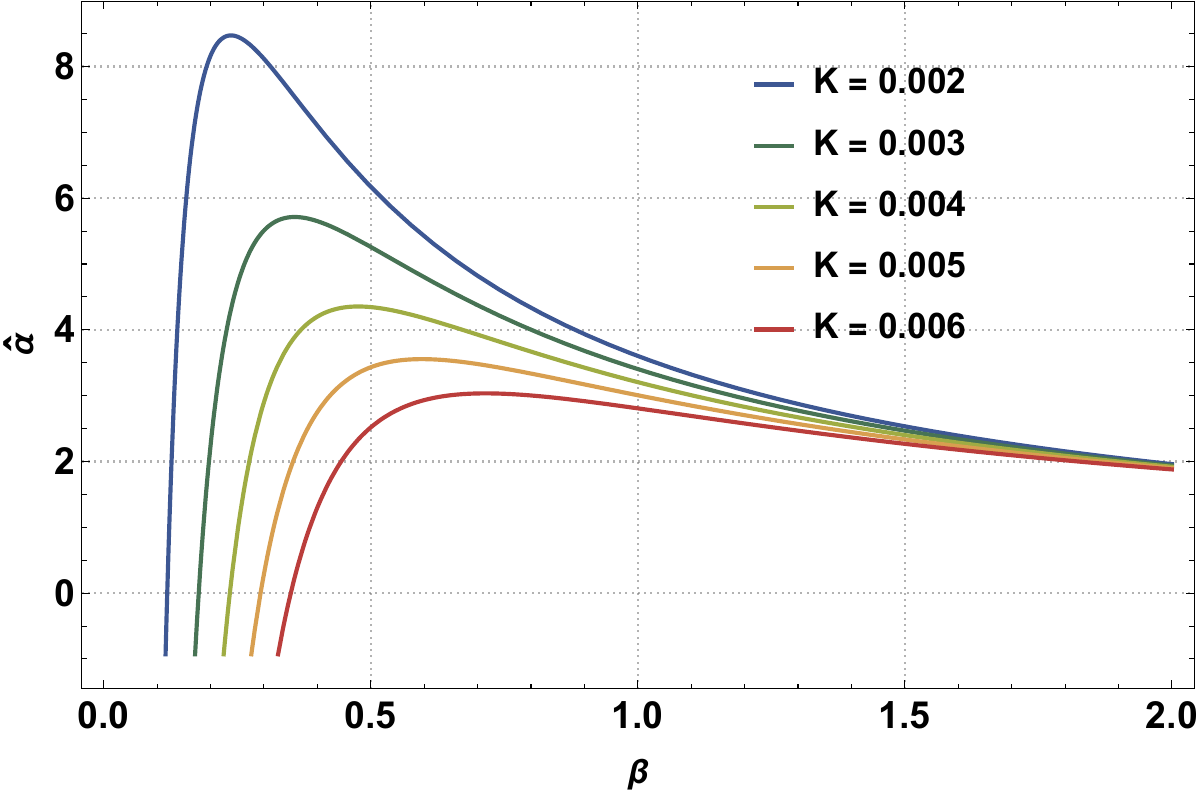}\\
(i) Varying $K$ ($e = 5$)\\[2mm]
\includegraphics[width=0.8\linewidth]{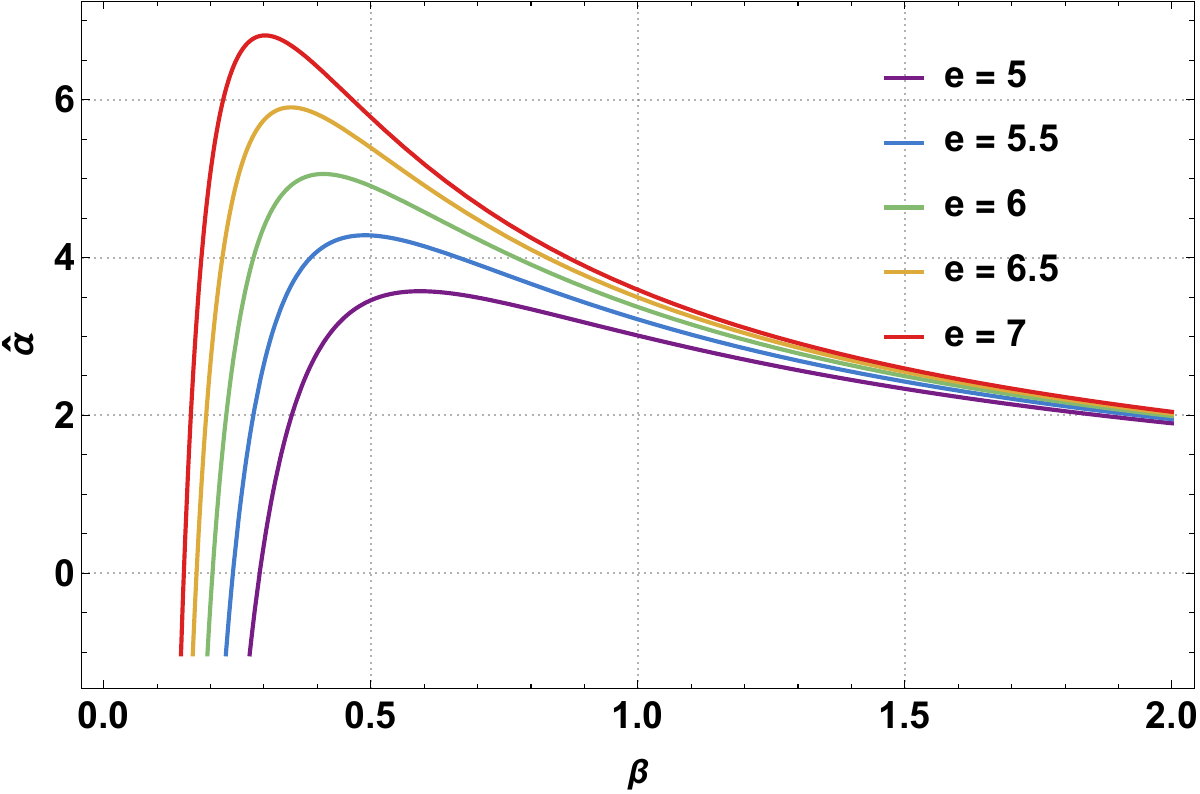}\\
(ii) Varying $e$ ($K = F_\pi^2/4$)
\caption{Deflection angle $\hat{\alpha}$ versus impact parameter $\beta$ for varying $K$ (upper) and $e$ (lower).}
\label{fig:deflection}
\end{figure}

Figure~\ref{fig:deflection} illustrates the deflection angle behavior predicted by Eq.~(\ref{ee7}). Panel~(i) demonstrates that increasing $K$ enhances both the constant offset $4\pi^2 K$ and the peak amplitude, consistent with the enlarged shadow radii observed in Table~\ref{tab:shadow} for larger $K$ values. Panel~(ii) shows that larger $e$ (corresponding to smaller $\lambda$) reduces the third term correction and shifts the peak toward smaller impact parameters. The deflection angle extremum, obtained by setting $\partial_\beta \hat{\alpha} = 0$, occurs at
\begin{equation}
\beta_{\rm peak} = \frac{3\pi^2 K\lambda}{2M},\label{ee9}
\end{equation}
with peak value
\begin{equation}
\hat{\alpha}_{\rm peak} = 4\pi^2 K + \frac{4M^2}{3\pi^2 K\lambda}.\label{ee10}
\end{equation}

\setlength{\tabcolsep}{10pt}
\renewcommand{\arraystretch}{1.5}
\begin{table}[ht!]
\centering
\begin{tabular}{|c|c|c|}
\hline
$K$ & $\beta_{\rm peak}$ & $\hat{\alpha}_{\rm peak}$ \\
\hline
0.002 & 0.2383 & 8.4722 \\
0.003 & 0.3574 & 5.7139 \\
0.004 & 0.4766 & 4.3545 \\
0.005 & 0.5957 & 3.5547 \\
0.006 & 0.7149 & 3.0346 \\
\hline
\end{tabular}
\caption{Values of $\beta_{\rm peak}$ and $\hat{\alpha}_{\rm peak}$ for varying $K$ with $e = 5$ ($\lambda = 8.0479$), $M = 1$.}
\label{tab:peak-K}
\end{table}

\vspace{1em}

\begin{table}[t!]
\centering
\begin{tabular}{|c|c|c|c|}
\hline
$e$ & $\lambda$ & $\beta_{\rm peak}$ & $\hat{\alpha}_{\rm peak}$ \\
\hline
5.0 & 8.0479 & 0.5922 & 3.5736 \\
5.5 & 6.6511 & 0.4894 & 4.2828 \\
6.0 & 5.5888 & 0.4112 & 5.0596 \\
6.5 & 4.7621 & 0.3504 & 5.9040 \\
7.0 & 4.1061 & 0.3021 & 6.8159 \\
\hline
\end{tabular}
\caption{Values of $\beta_{\rm peak}$ and $\hat{\alpha}_{\rm peak}$ for varying $e$ with $K = F_\pi^2/4$, $M = 1$.}
\label{tab:peak-e}
\end{table}

Tables~\ref{tab:peak-K} and~\ref{tab:peak-e} present numerical values of the peak quantities from Eqs.~(\ref{ee9})--(\ref{ee10}). Table~\ref{tab:peak-K} shows that increasing $K$ at fixed $\lambda$ shifts $\beta_{\rm peak}$ outward while reducing $\hat{\alpha}_{\rm peak}$, reflecting the enhanced Skyrme contribution that spreads the deflection over larger impact parameters. Table~\ref{tab:peak-e} reveals the opposite trend: as $e$ increases (decreasing $\lambda$), both $\beta_{\rm peak}$ and the product $K\lambda$ decrease, causing the peak to migrate inward while the peak amplitude $\hat{\alpha}_{\rm peak}$ grows due to the $1/(K\lambda)$ dependence in Eq.~(\ref{ee10}). These complementary behaviors provide distinct observational signatures: measuring both $\beta_{\rm peak}$ and $\hat{\alpha}_{\rm peak}$ would constrain the product $K\lambda$ and the ratio $M^2/(K\lambda)$ independently, potentially allowing extraction of the Skyrme parameters from deflection observations. Combined with shadow measurements from Sec.~II\,B, this offers complementary probes of the underlying Skyrme field configuration.

\begin{figure*}
    \centering
  \includegraphics[width=0.7\linewidth]{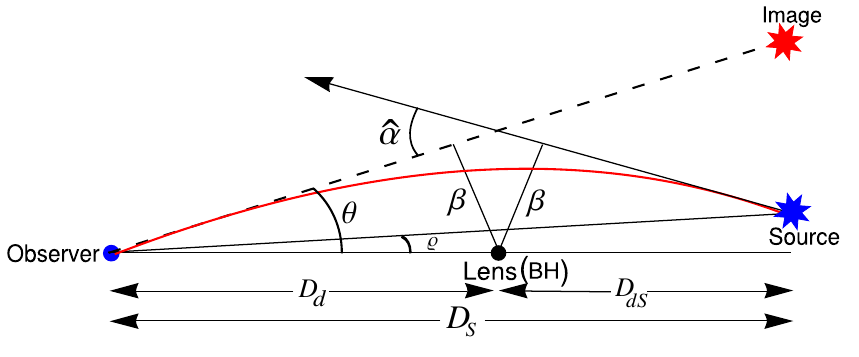}
    \caption{The lens equation geometry.}
    \label{fig:lens}
\end{figure*}

\section{Lens Equation and Magnification}\label{sec:lens}

The deflection angle analysis in Sec.~II\,D provides the foundation for studying gravitational lensing in the Skyrmion BH spacetime. Adopting the thin-lens formalism~\cite{Schneider1992,Virbhadra2000}, the lens equation relating angular source position $\varrho$ to image position $\theta$ takes the form
\begin{equation}
\varrho = \theta - m_1 - \frac{m_2}{\theta} + \frac{m_3}{\theta^2},\label{lens1}
\end{equation}
where $\beta \approx D_d \theta$ and the lensing coefficients are
\begin{equation}
m_1 = \frac{4\pi^2 K D_{dS}}{D_S},\quad m_2 = \frac{4M D_{dS}}{D_S D_d},\quad m_3 = \frac{3\pi^2 K\lambda D_{dS}}{D_S}.\label{lens2}
\end{equation}
Here $D_d$, $D_S$, and $D_{dS}$ denote the observer-lens, observer-source, and lens-source angular diameter distances. Each coefficient corresponds directly to a term in Eq.~(\ref{ee7}): $m_1$ encodes the Skyrme coupling contribution analogous to the global monopole deficit angle discussed after Eq.~(\ref{cc5}), $m_2$ defines the Einstein angle $\theta_E = \sqrt{m_2}$ governing standard Schwarzschild lensing, and $m_3$ represents the quartic Skyrme correction. Rearranging yields the cubic $\theta^3 - (m_1 + \varrho)\theta^2 - m_2\theta + m_3 = 0$, whose discriminant $\Delta = \Delta_2^2 - 4\Delta_1\Delta_3$ with $\Delta_1 = (\varrho + m_1)^2 + 3m_2$, $\Delta_2 = -9m_3 + (\varrho + m_1)m_2$, and $\Delta_3 = m_2^2 + (\varrho + m_1)m_3$ determines image multiplicity: three real images form when $\Delta < 0$ with $\Delta_1 > 0$, while $\Delta > 0$ yields one real image.

In the weak-field regime ($m_3 \ll m_2$), image positions reduce to $\theta_\pm = \frac{1}{2}[(m_1 + \varrho) \pm \sqrt{(m_1 + \varrho)^2 + 4m_2}]$, with magnification components~\cite{Ahmed:2025vww}
\begin{equation}
\mu_{\rm tan} = \left|1 - \frac{\theta_E^2}{\theta^2}\right|^{-1},\quad \mu_{\rm rad} = \left|1 + \frac{\theta_E^2}{\theta^2}\right|^{-1}.\label{lens3}
\end{equation}
The total magnification $\mu_{\rm tot} = \mu_{\rm tan} \times \mu_{\rm rad}$ diverges at $\theta = \pm\theta_E$, corresponding to Einstein ring formation when $\varrho = 0$.

\begin{figure}[ht!]
\centering
\includegraphics[width=0.95\linewidth]{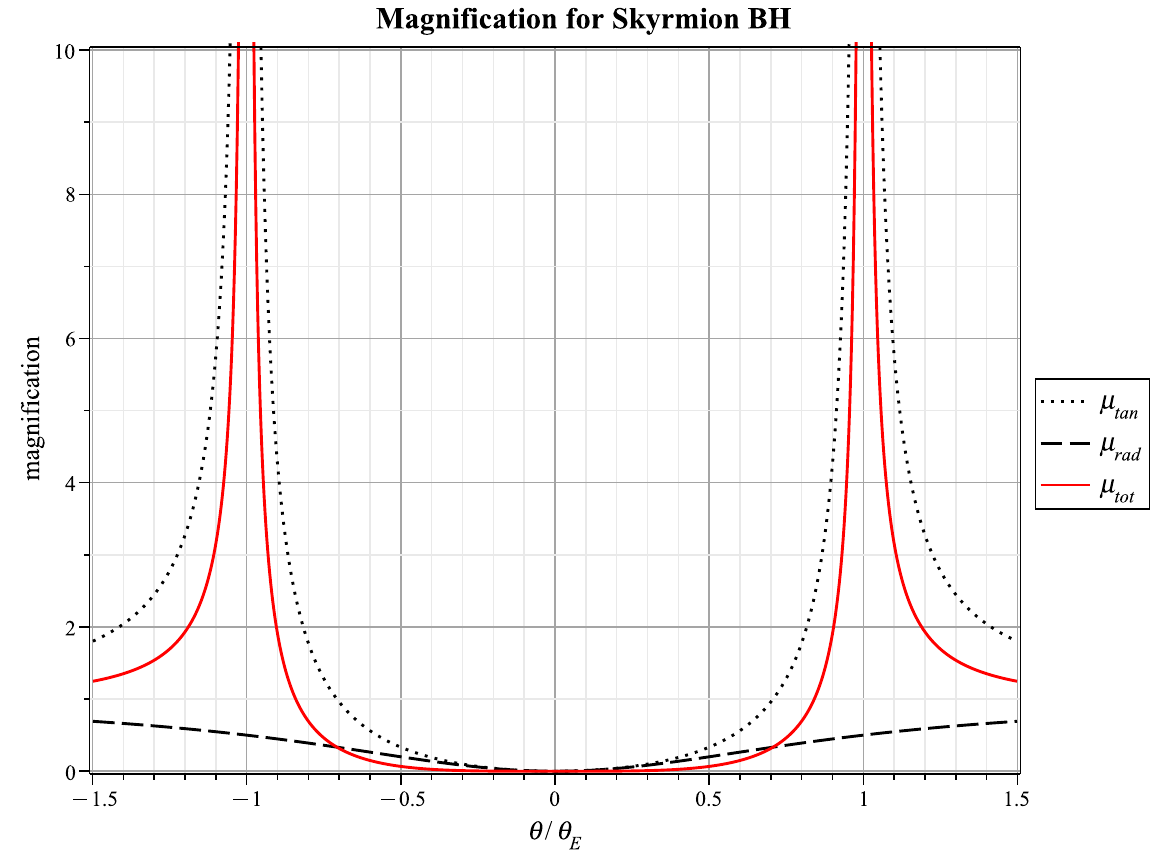}
\caption{Magnification components $\mu_{\rm tan}$ (dotted), $\mu_{\rm rad}$ (dashed), and $\mu_{\rm tot}$ (solid) versus $\theta/\theta_E$, showing divergence at the Einstein ring position.}
\label{fig:magnification}
\end{figure}

Figure~\ref{fig:magnification} displays the magnification components as functions of normalized angular position $\theta/\theta_E$. The tangential magnification $\mu_{\rm tan}$ diverges at $\theta = \pm\theta_E$ where caustic crossing occurs, while the radial component $\mu_{\rm rad}$ remains bounded throughout. The total magnification $\mu_{\rm tot}$ exhibits the characteristic point-lens profile with symmetric divergences, a feature preserved regardless of the Skyrme parameter values. The lensing coefficients~(\ref{lens2}) inherit the parameter dependence from the deflection angle, so the peak structure quantified in Tables~\ref{tab:peak-K}--\ref{tab:peak-e} translates into observable lensing signatures that complement the shadow constraints discussed in Sec.~II\,B.

\section{Sparsity of Hawking Radiation}

We now quantify Hawking radiation sparsity for the Skyrmion BH. Although BHs radiate thermally with temperature set by surface gravity, resembling classical black bodies, Hawking flux is temporally discrete with well-separated quantum emissions rather than continuous streaming. Sparsity measures the average time between consecutive quantum emissions normalized by the energy-determined time scale, comparing the squared thermal wavelength $\lambda_t=2\pi/T$ with effective horizon area $\mathcal{A}_{\rm eff}$~\cite{Page1976a,Page1976b,PagePhD1976,Page1977,Gray2016}:
\begin{equation}
    \psi =\frac{\mathcal{C}}{\tilde{g}}\left(\frac{\lambda_t^2}{\mathcal{A}_{\rm eff}}\right),\label{ff1}
\end{equation}
where $\mathcal{C}=1$ is a dimensionless constant, $\tilde{g}$ is the spin-degeneracy factor, $\mathcal{A}_{\rm eff}=\frac{27}{4}\mathcal{A}_{\rm BH}$ is the effective radiating area, and $T$ is the Hawking temperature. For Schwarzschild BHs, $T=1/(4\pi r_s)$ and $\lambda_t=8\pi^2 r_s$ with $r_h=r_s$. For massless spin-1 particles ($\tilde{g}=1$), $\psi_{\rm Sch.} = 64\pi^3/27 \approx 73.5$. Classical black bodies with equal surface area have $\psi \ll 1$, demonstrating extraordinarily sparse BH emission.

The Hawking temperature follows from surface gravity $\kappa$~\cite{Yu1994}:
\begin{equation}
    \kappa=-\frac{1}{2}\lim_{r \to r_h} \frac{\mathcal{D}\,\partial_r g_{tt}}{\sqrt{-g_{tt}\,g_{rr}}}=\frac{\mathcal{D}f'(r_h)}{2}=\frac{\mathcal{D}}{2 r_h}\left(1-8\pi K-\frac{4\pi K \lambda}{r_h^2}\right),\label{ff2}
\end{equation}
where
\begin{equation}
    \mathcal{D}=1/\sqrt{-\lim_{r \to \infty} g_{tt}}=(1-\eta^2)^{-1/2}.\label{ff3}
\end{equation}
For asymptotically flat spacetime where $\lim_{r \to \infty} f(r) \to 1$, $\mathcal{D}=1$.

The Hawking temperature and thermal wavelength are
\begin{align}
    T&=\frac{\kappa}{2\pi}=\frac{(1-8\pi K)^{-1/2}}{4\pi r_h}\left(1-8\pi K-\frac{4\pi K \lambda}{r_h^2}\right),\nonumber\\
    \lambda_t&=8\pi^2 r_h (1-8\pi K)^{1/2} \left(1-8\pi K-\frac{4\pi K \lambda}{r_h^2}\right)^{-1}.\label{ff4}
\end{align}

The sparsity parameter becomes
\begin{equation}
\psi=(1-8\pi K)\left(1-8\pi K-\frac{4\pi K \lambda}{r_h^2}\right)^{-2}\psi_{\rm Sch.},\label{ff5} 
\end{equation}
using $\mathcal{A}_{\rm eff}=27\pi r^2_h$ with horizon radius
\begin{equation}
    r_h=\frac{M+\sqrt{M^2-4\pi K \lambda (1-8\pi K)}}{1-8\pi K}.\label{ff6}
\end{equation}

\begin{figure}[ht!]
    \centering
    \includegraphics[width=0.8\linewidth]{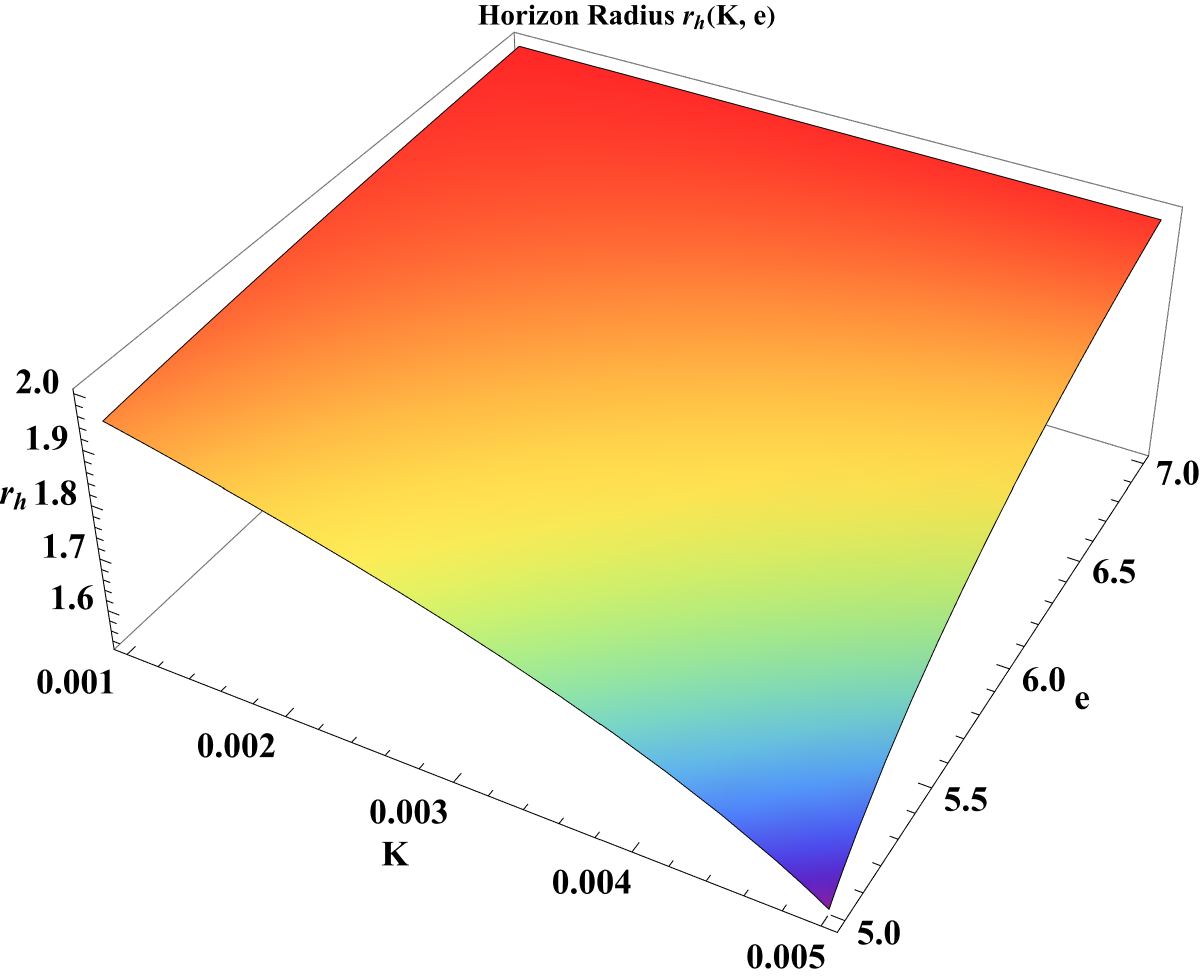}
    \caption{Three-dimensional surface of the horizon radius $r_h$ as a function of $K$ and $e$ with $M=1$.}    
    \label{fig:horizon}
\end{figure}

Expressions~(\ref{ff5}) and~(\ref{ff6}) show that $K$ and $\lambda$ affect both horizon radius and Hawking radiation sparsity. Figure~\ref{fig:horizon} displays the horizon radius as a function of $K$ and $e$, showing that $r_h$ increases with both $K$ and decreasing $e$ (increasing $\lambda$). Table~\ref{tab:shadow} lists horizon radius values, while Tables~\ref{tab:sparsity-1}--\ref{tab:sparsity-2} provide dimensionless sparsity parameters for various $K$ and $e$ values.

\begin{table*}[tbhp]
\centering
\setlength{\tabcolsep}{4pt} 
\renewcommand{\arraystretch}{1.2}
\begin{tabular}{|c|ccccccccccc|}
\hline
$e$ & 5.0 & 5.2 & 5.4 & 5.6 & 5.8 & 6.0 & 6.2 & 6.4 & 6.6 & 6.8 & 7.0 \\
\hline
$\psi / \psi_{\rm Sch}$ & 2.2967 & 1.9682 & 1.7886 & 1.6729 & 1.5915 & 1.5307 & 1.4835 & 1.4458 & 1.4149 & 1.3891 & 1.3673 \\
\hline
\end{tabular}
\caption{Sparsity ratio $\psi / \psi_{\rm Sch}$ for selected $e$ values with $M=1$ and $F_{\pi}=0.141$.}
\label{tab:sparsity-1}
\end{table*}

Table~\ref{tab:sparsity-1} shows that the sparsity ratio $\psi/\psi_{\rm Sch}$ decreases monotonically from 2.30 to 1.37 as $e$ increases from 5 to 7, indicating that larger $e$ values (smaller $\lambda$) bring the radiation sparsity closer to the Schwarzschild value.

\begin{table*}[tbhp]
\centering
\setlength{\tabcolsep}{4pt} 
\renewcommand{\arraystretch}{1.2}
\begin{tabular}{|c|ccccccccc|}
\hline
$K$ & 0.001 & 0.0015 & 0.002 & 0.0025 & 0.003 & 0.0035 & 0.004 & 0.0045 & 0.005 \\
\hline
$\psi / \psi_{\rm Sch}$ & 1.09104 & 1.14216 & 1.19760 & 1.25788 & 1.32363 & 1.39554 & 1.47446 & 1.56136 & 1.65743 \\
\hline
\end{tabular}
\caption{Sparsity ratio $\psi / \psi_{\rm Sch}$ for varying $K$ with $e=5$ and $M=1$.}
\label{tab:sparsity-2}
\end{table*}

Table~\ref{tab:sparsity-2} demonstrates that increasing $K$ from 0.001 to 0.005 raises $\psi/\psi_{\rm Sch}$ from 1.09 to 1.66, showing enhanced radiation sparsity with stronger Skyrme coupling.

\section{Energy Emission Rate}

Particle creation and annihilation near the horizon, driven by quantum fluctuations, produce emission energy. BH evaporation results from positive-energy particles tunneling out of the horizon region as Hawking radiation. For distant observers, the BH shadow corresponds to the high-energy absorption cross section approaching a limiting value $\sigma_{\rm lim}$~\cite{MTW1973,Mashhoon1973,Sanchez1978}.

Null geodesics govern high-energy quantum capture, so the limiting absorption cross-section can be estimated from the shadow~\cite{WeiLiu2013}:
\begin{equation}
\sigma_{\rm lim}\approx \pi R_{\rm sh}^{2},
\label{eer1}
\end{equation}
where $R_{\rm sh}$ is given in Eq.~(\ref{cc7}).

\begin{figure}[ht!]
    \centering
    \includegraphics[width=0.8\linewidth]{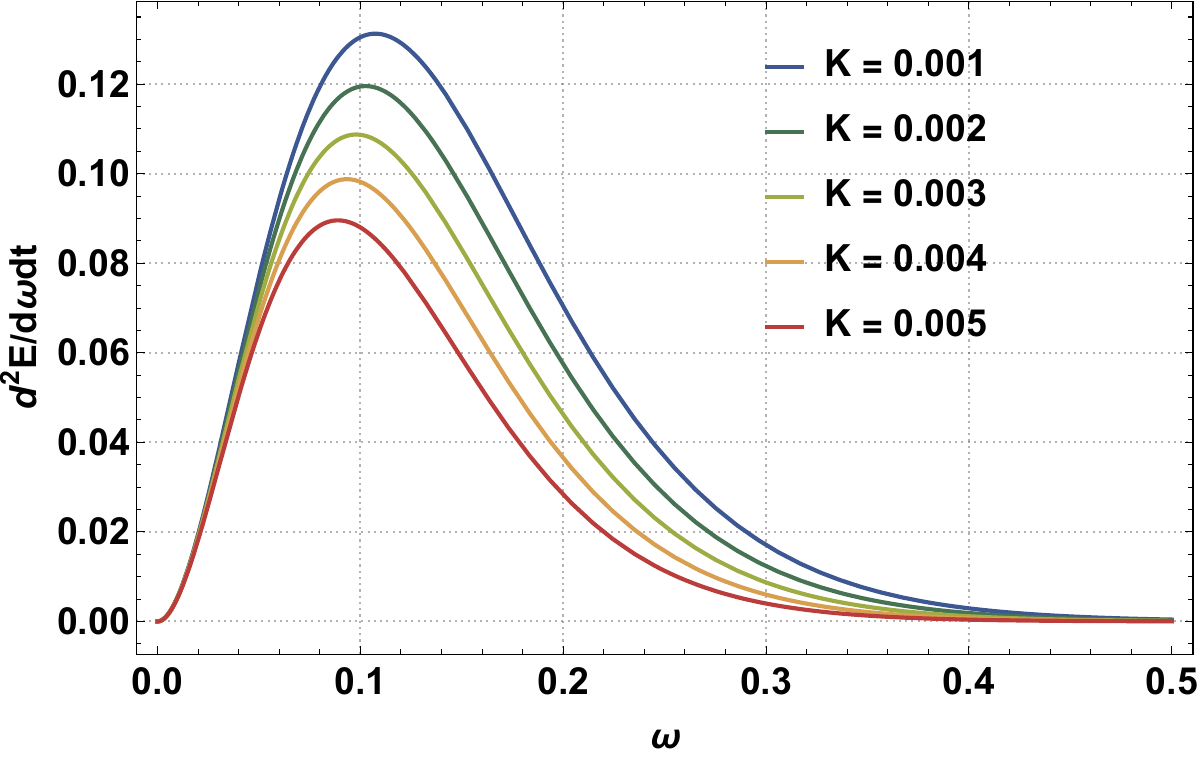}\\
    (i) $e=5$\\
    \includegraphics[width=0.8\linewidth]{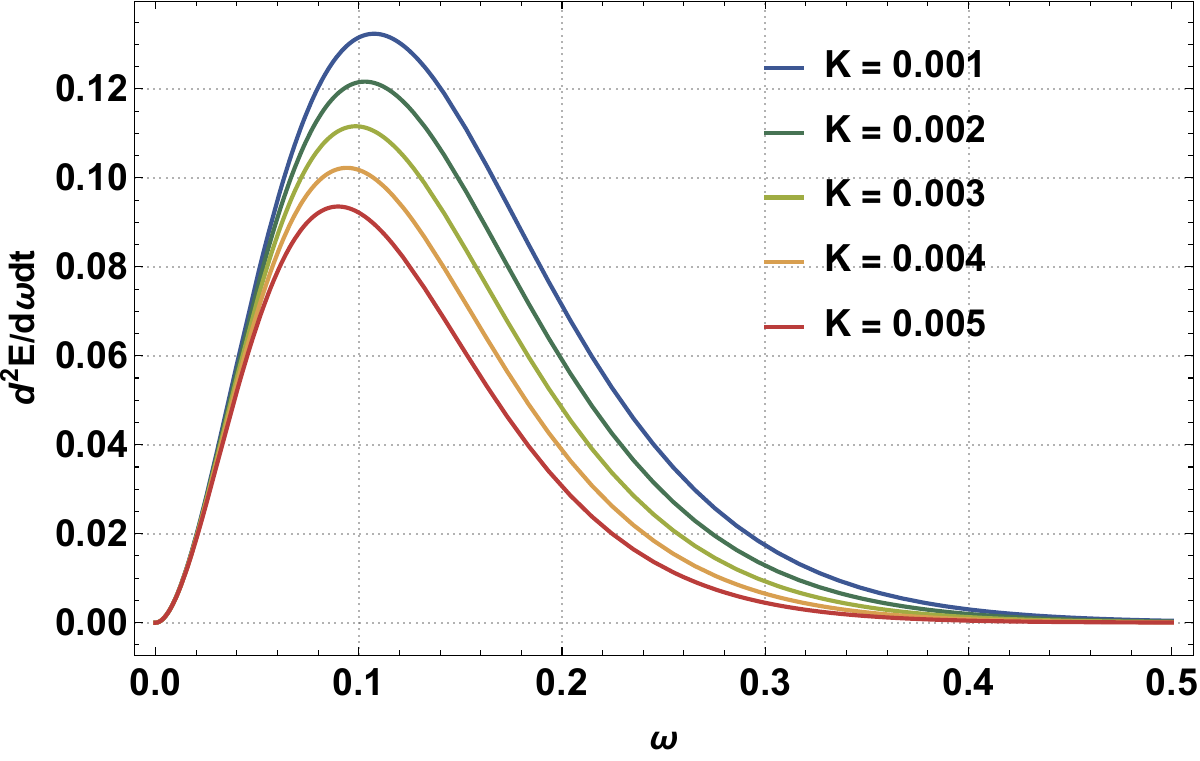}\\
    (ii) $e=6$ \\
    \includegraphics[width=0.8\linewidth]{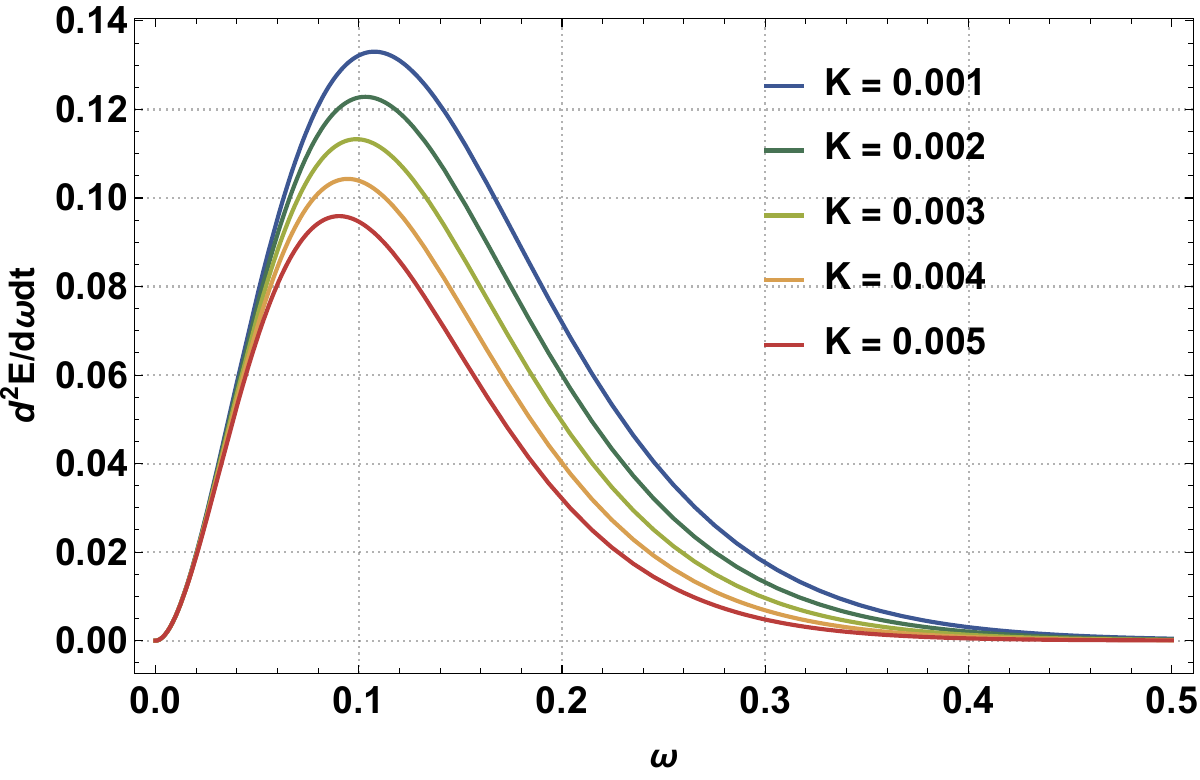}\\
    (iii) $e=7$ 
    \caption{Energy emission rate versus frequency $\omega$ for various $K$ with $e=5, 6, 7$. Here $M=1$.}
    \label{fig:energy-rate-1}
\end{figure}

The spectral energy emission rate is~\cite{WeiLiu2013,Hendi2023}
\begin{equation}
\frac{d^{2}\mathbb{E}}{d\omega dt}
=
\frac{2\pi^{2}\sigma_{\rm lim}}{e^{\omega/T}-1}\,\omega^{3},
\label{eer2}
\end{equation}
where $\omega$ is emission frequency and $T$ is the Hawking temperature.

Substituting Eqs.~(\ref{ff4}) and (\ref{eer1}) yields
\begin{equation}
\frac{d^{2}\mathbb{E}}{d\omega dt}
=
\frac{2\pi^{3}R_{\rm sh}^{2}\,\omega^{3}}{e^{\omega/T_H}-1}.
\label{eer3}
\end{equation}

Using Eq.~(\ref{ff4}) gives explicitly
\begin{equation}
\frac{d^{2}\mathbb{E}}{d\omega dt}
=
\frac{2\pi^{3} R_{\rm sh}^{2}\,\omega^{3}}
{\exp\!\left\{4\pi r_h \omega \sqrt{1-8\pi K}\left(1-8\pi K-\frac{4\pi K \lambda}{r_h^{2}} \right)^{-1}\right\}-1},
\label{eer4}
\end{equation}
where $r_h$ is given by Eq.~(\ref{ff6}).

Equation~(\ref{eer4}) shows that both $\lambda$ and $K$ control the energy emission rate.

Figure~\ref{fig:energy-rate-1} displays the energy emission rate versus frequency $\omega$ for three $e$ values. At fixed $e$, the emission rate peak decreases as $K$ increases, reflecting reduced effective temperature and enlarged shadow radius. Comparing panels (i)--(iii), larger $e$ values yield higher peak emission rates for the same $K$ range.

\begin{figure}[ht!]
    \centering
    \includegraphics[width=0.8\linewidth]{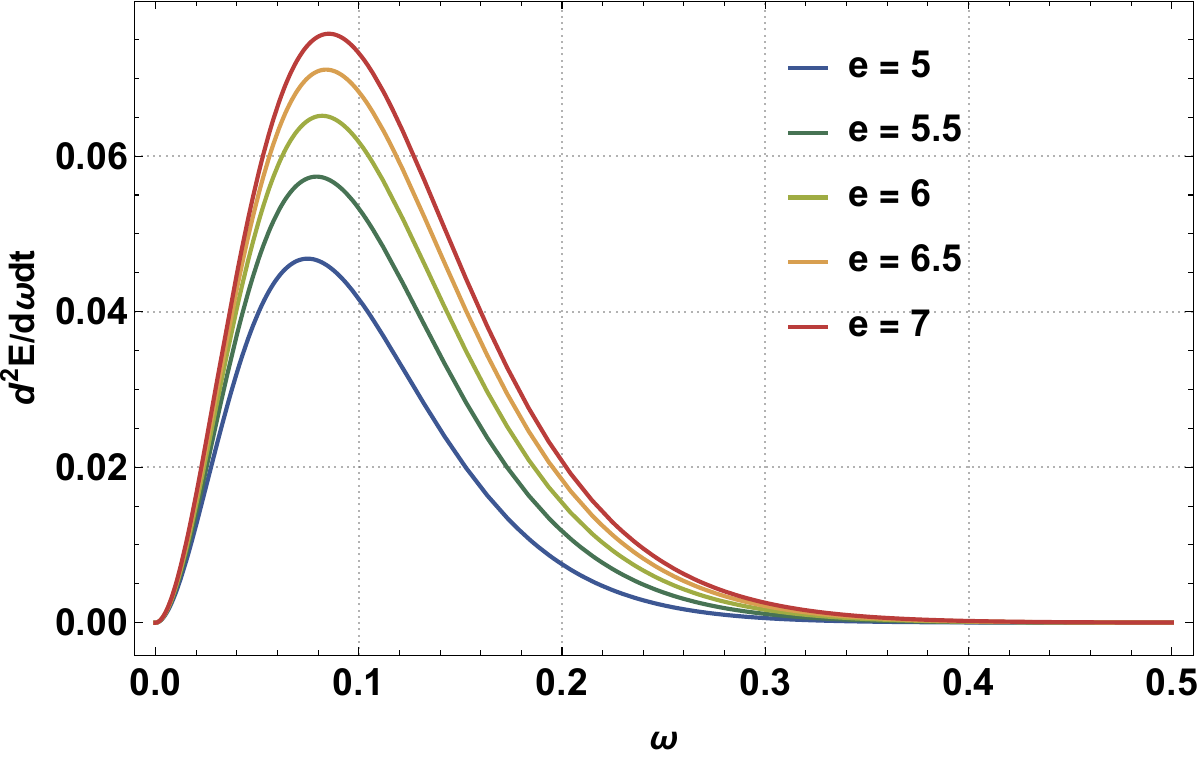}
    \caption{Energy emission rate versus frequency $\omega$ for various $e$ with $F_{\pi}=0.141$ and $M=1$.}
    \label{fig:energy-rate-2}
\end{figure}

Figure~\ref{fig:energy-rate-2} shows the emission rate for varying $e$ at fixed $F_\pi$. The peak increases with $e$, consistent with smaller $\lambda$ values producing higher effective temperatures. These emission characteristics may offer insights for BH detection strategies.

\section{Conclusion}\label{sec:5}
 
We have examined photon dynamics, gravitational lensing, and radiative properties of Skyrmion BHs arising from the Einstein-Skyrme system. The Skyrme coupling constant $K = F_{\pi}^2/4$ and quartic parameter $\lambda = 4/(e^2 F_{\pi}^2)$ modify the metric function similarly to the Reissner-Nordstr\"om charge term, though here the coefficient derives from hadron physics rather than being a free integration constant.
 
The photon sphere radius~(\ref{cc3}) and shadow radius~(\ref{cc7}) both grow with increasing $K$ while shrinking as $e$ increases, as confirmed numerically in Table~\ref{tab:shadow}. The two-dimensional intensity maps of Fig.~\ref{fig:shadow-grid} visualise the same trend directly on the $(K,e)$ plane. The constraint~(\ref{cc4}) ensures photon sphere existence and is comfortably satisfied for phenomenologically relevant parameters. The asymptotic deficit angle~(\ref{cc5}) produces a constant offset $4\pi^2 K$ in the deflection angle, distinguishing Skyrmion lensing from Schwarzschild predictions.
 
The weak-field deflection angle~(\ref{ee7}) exhibits a three-term structure: a Skyrme-induced constant, the standard $4M/\beta$ contribution, and a quartic correction scaling as $1/\beta^2$. This structure produces a deflection maximum at finite impact parameter given by Eqs.~(\ref{ee9})--(\ref{ee10}). Tables~\ref{tab:peak-K} and~\ref{tab:peak-e} show that increasing $K$ shifts $\beta_{\rm peak}$ outward while reducing $\hat{\alpha}_{\rm peak}$, whereas increasing $e$ produces the opposite trend. Combined measurements of these peak quantities would constrain the Skyrme parameters independently, complementing shadow observations.
 
The lens equation~(\ref{lens1}) translates deflection results into observable signatures through coefficients~(\ref{lens2}), with $m_1$ encoding the global monopole-like shift, $m_2$ defining the Einstein angle, and $m_3$ representing the quartic correction. Image multiplicity depends on a cubic discriminant that can yield three-image configurations absent in standard Schwarzschild lensing.
 
Hawking radiation sparsity~(\ref{ff5}) exceeds the Schwarzschild value throughout the relevant parameter range, with $\psi/\psi_{\rm Sch}$ reaching 2.30 at $e=5$ and 1.66 at $K=0.005$ as shown in Tables~\ref{tab:sparsity-1}--\ref{tab:sparsity-2}. The energy emission rate~(\ref{eer4}) depends on both shadow radius and Hawking temperature; increasing $K$ suppresses peak emission while increasing $e$ enhances it, as illustrated in Figs.~\ref{fig:energy-rate-1}--\ref{fig:energy-rate-2}.
 
These results demonstrate that Skyrmion BHs possess distinctive optical and radiative signatures: enlarged photon spheres and shadows, deflection angles with finite-impact-parameter peaks, enhanced radiation sparsity, and modified emission spectra. While current observational precision cannot yet probe these effects, next-generation very long baseline interferometry (VLBI) arrays and gravitational lensing surveys may eventually test these predictions, connecting BH physics with hadronic field theory.
 
\scriptsize

\section*{Acknowledgments}
 
F.A. acknowledges the Inter University Centre for Astronomy and Astrophysics (IUCAA), Pune, India for granting visiting associateship. \.{I}.~S. expresses gratitude to T\"{U}B\.{I}TAK, ANKOS, and SCOAP3 for their academic support. He also acknowledges COST Actions CA22113, CA21106, CA21136, CA23130, and CA23115 for their contributions to networking.

\scriptsize
 
\bibliographystyle{apsrev4-2}

\bibliography{ref.bib}
 
\end{document}